\documentclass[aps,prd,twocolumn,preprintnumbers,nofootinbib,superscriptaddress, showkeys]{revtex4-2}

\usepackage{lineno}
\usepackage[usenames]{color}
\usepackage{changepage} 
\usepackage{amssymb}
\usepackage{amsmath}
\usepackage{rotating}
\usepackage{comment}
\usepackage{graphicx}
\usepackage{caption}
\usepackage{subcaption}
\usepackage[dvipsnames]{xcolor}
\usepackage{hyperref}
\hypersetup{
    colorlinks=true,
    linkcolor=blue,
    citecolor=blue,
    urlcolor=blue
}
\usepackage{soul}
\usepackage{ulem}

\begin{document}



\title{Sgr A* as a Galactic PeVatron: Multimessenger Signatures of the Magnetic Penrose Process}


\author{Marina Cermeño}
\affiliation{Departamento de FAIAN, ETSIN, Universidad Politécnica de Madrid, Av. de la Memoria~4, E-28040, Madrid, Spain}
\affiliation{Instituto de F\'isica Te\'orica UAM-CSIC, C/ Nicol\'as Cabrera, 13-15, 28049 Madrid, Spain}

\author{Pedro De La Torre Luque}
\affiliation{Instituto de F\'isica Te\'orica UAM-CSIC, C/ Nicol\'as Cabrera, 13-15, 28049 Madrid, Spain}
\affiliation{Departamento de F\'isica Te\'orica, M-15, Universidad Aut\'onoma de Madrid, E-28049 Madrid, Spain}

\author{Cristina Fernández-Suárez}
\email{cristina.fernandezs01@estudiante.uam.es}
\thanks{Corresponding author}
\affiliation{Instituto de F\'isica Te\'orica UAM-CSIC, C/ Nicol\'as Cabrera, 13-15, 28049 Madrid, Spain}
\affiliation{Departamento de F\'isica Te\'orica, M-15, Universidad Aut\'onoma de Madrid, E-28049 Madrid, Spain}

\author{Viviana Gammaldi}
\affiliation{Department of Information Technology, Escuela Polit\'ecnica Superior, Universidad San Pablo-CEU, Campus Montepr\'incipe, Boadilla del Monte, Madrid, 28668, Spain}

\author{Enrique Mier-Alonso}
\email{enrique.mier@estudiante.uam.es}
\thanks{Corresponding author}
\affiliation{Instituto de F\'isica Te\'orica UAM-CSIC, C/ Nicol\'as Cabrera, 13-15, 28049 Madrid, Spain}
\affiliation{Departamento de F\'isica Te\'orica, M-15, Universidad Aut\'onoma de Madrid, E-28049 Madrid, Spain}

\author{María J. Rodríguez}
\affiliation{Department of Physics, Utah State University, 4415 Old Main Hill Road, Logan, UT 84322, USA}
\affiliation{Instituto de F\'isica Te\'orica UAM-CSIC, C/ Nicol\'as Cabrera, 13-15, 28049 Madrid, Spain}

\author{Miguel Á. Sánchez-Conde}
\affiliation{Instituto de F\'isica Te\'orica UAM-CSIC, C/ Nicol\'as Cabrera, 13-15, 28049 Madrid, Spain}
\affiliation{Departamento de F\'isica Te\'orica, M-15, Universidad Aut\'onoma de Madrid, E-28049 Madrid, Spain}

\author{Jaume Zuriaga-Puig}
\affiliation{Instituto de F\'isica Te\'orica UAM-CSIC, C/ Nicol\'as Cabrera, 13-15, 28049 Madrid, Spain}
\affiliation{Departamento de F\'isica Te\'orica, M-15, Universidad Aut\'onoma de Madrid, E-28049 Madrid, Spain}





\begin{abstract}

The magnetic Penrose process (MPP) is one of the most efficient mechanisms for extracting rotational energy from a magnetized Kerr black hole (BH), enabling the acceleration of charged particles up to very-high energies. In this work, we investigate this process in Sagittarius A* (Sgr~A*), the supermassive black hole (SMBH) at the Galactic center (GC), focusing on its ultra-high-efficiency regime, in which neutrons undergo beta decay inside the BH ergosphere. 
As a key novelty, we compute the neutron production spectrum in the accretion flow directly from the underlying nuclear reaction kinematics and follow the neutron trajectories in the Kerr spacetime to determine the population that reaches the ergosphere and undergoes the MPP.  
From this population, we consistently derive the spectrum of accelerated protons and show that Sgr~A* can accelerate them up to PeV energies, strengthening its interpretation as a candidate Galactic PeVatron.
We further compute the gamma-ray and neutrino emission produced through hadronic interactions of the escaping protons in the Central Molecular Zone. The predicted gamma-ray fluxes exhibit distinctive spectral features that could provide an observational signature of the MPP 
and may constitute a non-negligible contribution to the very-high-energy emission detected by H.E.S.S. and HAWC.  
The associated neutrino fluxes remain below the diffuse Galactic component inferred by IceCube, although they may still contribute to the high-energy emission from the GC. 
Remarkably, the predicted signals lie within the projected sensitivity of SWGO for all the scenarios considered in this work and, for some models, only a factor of a few below the nominal CTAO sensitivity, while future neutrino observatories including KM3NeT/ARCA and IceCube-Gen2 will provide complementary tests of this 
scenario. 
Our results establish the MPP as a realistic and observationally testable mechanism for extracting BH rotational energy, while providing a direct connection between horizon-scale physics and high-energy multi-messenger observations that can be naturally extended to other magnetized BHs.

\end{abstract}


\maketitle


\section{Introduction}

Supermassive black holes (SMBHs), with masses greater than $10^5 \; \rm M_{\odot}$, are among the largest energy reservoirs in the Universe. In particular, rotating (Kerr) black holes (BHs) can store up to $\sim 29\%$ of their mass in the form of extractable rotational energy that can be converted into energy of outgoing particles and fields~\cite{Christodoulou1970, Christodoulou1971, rueda2024kerrblackholeenergy}. 

A broad framework for extracting rotational energy from a Kerr BH is provided by the phenomenon of \textit{superradiance} (see, e.g., Ref.~\cite{Brito_2020} for a review), in which waves, fields, or particles interacting with the rotating BH or compact object can be amplified at the expense of its rotational energy. 
A pioneering particle-based realization of this mechanism was proposed by Roger Penrose in 1969, namely the Penrose process (PP;~\cite{Penrose:1969pc}). In this scenario, particles can extract rotational energy from the BH through a decay process occurring within the ergosphere, i.e., the region where spacetime is dragged and forced to co-rotate with the BH, allowing the existence of negative-energy orbits.
The underlying mechanism consists of a particle decaying within the ergosphere into two fragments, one of which attains negative energy as measured by an observer at infinity, while the other escapes with an energy greater than that of the initial particle, thereby extracting rotational energy from the BH. However, the original PP is characterized by a relatively low efficiency (defined in the standard way as the ratio between the extracted and input energies) and requires highly relativistic relative velocities — exceeding half the speed of light — between the decay products within the ergosphere~\cite{1972ApJ...178..347B, PhysRevD.10.1680}, 
which limits its astrophysical relevance in most realistic environments. 

This has motivated the development of modified versions of the original PP (e.g.,~\cite{1975Piran, Kolo__2021, Kokubu_2021, Turs2021_electric}), including 
the so-called \textit{magnetic Penrose process} (MPP;~\cite{1985ApJ...290...12W}). The latter corresponds to the magnetic version of the PP, where the BH is embedded in an external magnetic field. The interaction between charged particles and the electromagnetic field enables them to access negative-energy orbits without the need for extreme kinematic conditions, removing the requirement of large relative velocities and rendering the process astrophysically viable, with efficiencies that can exceed $100\%$~\cite{1985JApA....6...85B, 1986ApJ...307...38P, 1989PhR...183..137W}. Another well-known electromagnetic mechanism for energy extraction is the Blandford–Znajek (BZ) process~\cite{1977BZ}, which can be regarded as a high limit of the MPP, requiring strong magnetic fields, typically above $\sim 10^{4} \; \rm G$ for stellar mass BHs~\cite{Dadhich_2018} . In this sense, the MPP provides a more general framework, encompassing both low and high magnetic field regimes, and naturally connecting to the BZ mechanism in the appropriate limit~\cite{dadhich2012magneticpenroseprocessblanfordzanejk}.

In this work, we investigate electromagnetic energy extraction in the SMBH at the center of our Galaxy, Sagittarius A* (Sgr~A*). 
Given the relatively low magnetic field expected in this system ($\sim 30$–$560 \; \rm G$, depending on the distance to the center~\cite{2024ETH_VIII}), the BZ process is unlikely to operate efficiently. We therefore focus on the MPP as the relevant energy extraction mechanism in this context. Its efficiency can span a wide range — from low to ultra-high — depending on the magnetic field strength and the properties of the particles involved~\cite{Tursunov_2019}. In the absence of a magnetic field, the process reduces to the original PP, with a maximum efficiency of $\sim 21\%$ for an extremally rotating BH~\cite{Tursunov_2019}.
In the presence of a magnetic field, if all particles are charged, the efficiency can be moderately enhanced, particularly when the escaping particle carries a larger charge than the incident one. Nevertheless, in this regime the efficiency remains limited by the requirement of global plasma neutrality in the BH environment. 
Finally, in the ultra-high efficiency regime, the incident particle is neutral and decays into two charged particles of opposite charge. In this case, the efficiency can exceed $100 \%$ for relatively weak magnetic fields of order $\sim \rm mG$~\cite{Dadhich_2018}, or even exceed values of $\sim 10^{12}$ under favorable conditions~\cite{Tursunov_2020}. 

The ultra-high efficiency regime of the MPP is particularly appealing, as it naturally produces high-energy escaping particles and may therefore lead to observable signatures. To provide a quantitative estimate of this mechanism, we consider neutron beta decay, in which a neutron decays within the ergosphere of Sgr~A* into charged particles—an electron and a proton. According to our magnetic field configuration (see Section~\ref{subsec:wald}), the energetically favored outcome is that the electron falls into the BH on a negative-energy orbit, while the proton escapes to infinity with enhanced energy, thereby extracting rotational energy from Sgr~A*. This scenario is especially well motivated in the case of Sgr~A*, whose accretion flow can be described within the advection-dominated accretion flow (ADAF;~\cite{Narayan_1994, Narayan_1995a, Narayan_1995b}) framework. In this regime, the plasma reaches sufficiently high ion temperatures to enable the production of neutrons through nuclear reactions, thus providing a natural source of the neutral particles required for the process. 

Previous studies have demonstrated the relevance of the MPP for energy extraction from SMBHs and highlighted its potential as a source of ultra-high-energy cosmic rays (UHECRs; $>1 \, \rm EeV$; e.g.,~\cite{Tursunov_2019, Tursunov_2020, Turs2022}). However, to our knowledge, this is the first work to provide a detailed quantitative estimate of this process in Sgr~A*, connecting neutron production in the accretion flow and the subsequent MPP acceleration to derive the spectrum of escaping protons. 
Although Ref.~\cite{Oh_2024} also presents an estimate of the proton production rate associated with this mechanism, our treatment extends their analysis in several important aspects. First, instead of adopting a Maxwell-Boltzmann (MB) distribution for the neutrons, together with a thermally averaged reaction rate, we compute the neutron production spectrum directly from the differential cross-section, thus accounting for the energy dependence imposed by the reaction kinematics, rather than assuming that neutrons inherit the thermal distribution of the background plasma.
Second, we follow the propagation of neutrons in the Kerr spacetime, consistently accounting for the rotation of the BH when evaluating their survival probability before reaching the ergosphere, whereas Ref.~\cite{Oh_2024} considered Schwarzschild trajectories. 
Finally, rather than estimating the proton energies from the local conditions at the neutron decay location, we derive the escaping proton spectrum by consistently incorporating the energy gain provided by the MPP, leading to a fully self-consistent prediction of the accelerated proton population.

Furthermore, we investigate the role of the MPP in Sgr~A* as an efficient injector of high-energy cosmic rays (HECRs; $>1 \, \rm PeV$), showing that it can operate as a competitive particle acceleration mechanism. We use our results to predict the gamma-ray and neutrino emission arising from interactions of the escaping accelerated protons with the surrounding gas. We compare our predictions with the very-high-energy (VHE; $>100 \, \rm GeV$) diffuse emission observed by the High Energy Stereoscopic System (H.E.S.S.;~\cite{Hinton_2004}) in the Galactic center (GC) region~\cite{2018hess}, as well as with the central point-like source detected in the same region by H.E.S.S.~\cite{2016hess} and the High-Altitude Water Cherenkov (HAWC;~\cite{HAWC:2013kjc}) Observatory~\cite{albert2024observationgalacticcenterpevatron}. Additionally, we assess the detectability of the predicted signals by confronting our results with the sensitivity curves of current and future gamma-ray and neutrino observatories, thereby exploring the observational prospects of the MPP in Sgr~A*.

This paper is organized as follows. In Section~\ref{sec:disk_model}, we detail the modeling for the accretion disk of Sgr~A*, and introduce the different scenarios considered throughout this work, which are defined by our choice of BH parameters.
Section~\ref{sec:neutron} presents our calculation of the neutron production rate within the accretion flow. In Section~\ref{sec:MPP}, we outline the MPP and its application to Sgr~A*, and derive the resulting spectrum of accelerated protons for the different scenarios considered in this work. Section~\ref{sec:obs_signatures} investigates the high-energy implications of this mechanism, including the potential production of HECRs and the associated gamma-ray and neutrino spectra arising from interactions of MPP-accelerated protons, together with their observational prospects.
We summarize our main results and discuss the primary caveats of this study in Section~\ref{sec:conclusions}.

Throughout this paper, we will use geometrical units ($G = c = 1$), unless explicitly stated otherwise.

\section{Accretion disk modeling}
\label{sec:disk_model}
Due to the low luminosity and low accretion rate of Sgr~A*, its accretion flow is well described by an ADAF model~\cite{Narayan_1994, Narayan_1995a, Narayan_1995b}. Moreover, recent observations from the Event Horizon Telescope (EHT; \cite{2019aEHT}) suggest that the accretion flow around Sgr~A* may be in a magnetically arrested disk (MAD;~\cite{Narayan_2003_MAD, Igumenshchev_2003}) state~\cite{Event_Horizon_Telescope_Collaboration_2022_V}. This Section is devoted to presenting these models, as well as their application to the particular case of Sgr~A*.

\subsection{Advection-dominated accretion flow (ADAF)}
\label{subsec:adaf}
The ADAF model~\cite{Narayan_1994, Narayan_1995a, Narayan_1995b} describes a hot, low-luminosity, geometrically thick accretion flow around compact objects. In the ADAF solution, the gas is usually modeled as a two-temperature plasma, with the ion temperature typically reaching $T_{i} \sim 10^{12} \: \rm K$ near the center, while electrons are cooler, with a maximum temperature of $T_{e} \sim 10^{9}-10^{11} \: \rm K$. 
The energy produced by viscous dissipation is predominantly advected radially with the accretion flow instead of being radiated away. Since the gas retains most of this energy, the resulting pressure is high. As a result, the accretion flow becomes geometrically thick, with a vertical height ($H$) comparable to the radial size ($R$), i.e., $H/R \sim 1$. 
The high internal pressure provides partial radial support against gravity, causing the angular velocity of the gas to be sub-Keplerian. At the same time, the radial velocity is relatively large, which results in a short accretion timescale. The combination of large radial velocities and the large scale height reduces the gas density, making the medium optically thin and inefficient at radiative cooling.

The main properties of ADAFs are captured by the self-similar ADAF solution in the Newtonian limit presented in Refs.~\cite{Narayan_1994, Narayan_1995b}, in which the physical quantities describing the accretion flow (such as the vertical scale height ($H$), the radial velocity ($v_r$), the angular rotation frequency ($\Omega$), 
or the mass density ($\rho$)) can be expressed as power-law functions of the radial distance (Eqs.~\eqref{eq:H}--\eqref{eq:rho}):
\begin{equation}
    H=(2.5 c_{3})^{1/2} R \; \rm [cm],
    \label{eq:H}
\end{equation}
\begin{equation}
    v_r = -2.12 \times 10^{10} \alpha c_{1} \tilde{r}^{-1/2} \; \rm [cm \, s^{-1}],
    \label{eq:vr}
\end{equation}
\begin{equation}
    \Omega = 7.19 \times 10^{4} c_{2} m^{-1} \tilde{r}^{-3/2} \; \rm [s^{-1}],
    \label{eq:Omega}
\end{equation}
\begin{equation}
    \rho = 3.79 \times 10^{5} \alpha^{-1} c_{1}^{-1} c_{3}^{-1/2} m^{-1} \dot{m} \tilde{r}^{-3/2} \; \rm [g \, cm^{-3}].
    \label{eq:rho}
\end{equation}

The above relations are expressed in terms of the scaled disk radius, $\tilde{r}=R/R_{\rm S}$ with 
$R_{\rm S}=2M$ being the Schwarzschild radius, 
BH mass, $m=M/M_{\odot}$, and mass accretion rate, $\dot{m}=\dot{M}/\dot{M}_{\rm Edd}$, with the Eddington rate related to the Eddington luminosity via $\dot{M}_{\rm Edd}=L_{\rm Edd} / 0.1$ (i.e., $\dot{M}$ is the mass accretion rate at which a disk with radiative efficiency 0.1 would radiate at the Eddington luminosity).\footnote{The Eddington luminosity provides an idealized estimate of the maximum luminosity that an accreting BH can sustain. It is defined as $L_{\rm Edd} \simeq 4\pi G M c m_p / \sigma_T \approx 10^{38} (M/M_\odot)\ \mathrm{erg\ s^{-1}}$, where $m_p$ is the proton mass and $\sigma_{T}=6.62 \times 10^{-25} \; \rm cm^2$ is the Thomson cross-section.}

The constants $c_1$, $c_{2}$ and $c_{3}$ are defined as: 
\begin{widetext}

\begin{equation}
    c_{1}=\frac{(5+2\epsilon')}{3\alpha^{2}}g(\alpha, \epsilon'), \quad
    c_{2}=\left(\frac{2\epsilon'(5+2\epsilon')}{9\alpha^{2}}g(\alpha, \epsilon'), \right)^{1/2}, \quad
    c_{3}=\frac{2(5+2\epsilon')}{9\alpha^{2}}g(\alpha, \epsilon'),
    \label{eq:c1c2c3}
\end{equation}

\end{widetext}
with:
\begin{equation}
    \epsilon'=\frac{1}{f_{\rm adv}}\left(\frac{5/3 - \gamma}{\gamma -1} \right),
    \label{eq:eps'}
\end{equation}
\begin{equation}
    g(\alpha, \epsilon') = \left(1+ \frac{18\alpha^{2}}{(5+2\epsilon')^{2}} \right)^{1/2} -1 .
    \label{eq:g_function}
\end{equation}

All these relations depend on the standard viscosity parameter  $\alpha$~\cite{Shakura1973}, the ratio of the gas pressure to the total pressure of the accreting gas $\beta$, i.e., $P = P_{g} + P_{m}$, where $P_{g}$ and $P_{m}$ denote the gas and magnetic pressure, respectively), the fraction of viscously dissipated energy which is advected $f_{\rm adv}$, and the adiabatic index $\gamma$, defined as: 
\begin{equation}
    \gamma = \frac{32 - 24\beta - 3\beta^{2}}{24 - 21\beta}.
    \label{eq:gamma}
\end{equation}

Additionally, in the two-temperature plasma framework, the ion and electron temperatures are related as: 

\begin{equation}
    T_{i} + 1.08 T_{e} = 6.66\times 10^{12} \beta c_{3} \tilde{r}^{-1} \; \rm [K].
    \label{eq:temp}
\end{equation}

The ADAF solution only applies for low luminosities (and generally low mass accretion rates). In particular, the gas density would be low enough to allow a two-temperature plasma only if $L \leq 0.1 \, L_{\rm Edd}$~\cite{Narayan_1995b}. 
Sgr~A* exhibits an extremely low luminosity of $\lesssim 10^{36} \; \rm erg \, s^{-1}$~\cite{Event_Horizon_Telescope_Collaboration_2022}, i.e., $L \lesssim 10^{-8} \, L_{\rm Edd}$, 
indicating that its accretion flow is compatible with an ADAF model.  
For a given BH mass and mass accretion rate, together with a specific choice of the parameters entering the expressions for the radial profiles of the gas quantities, the local properties of the accretion flow can be fully determined. According to direct imaging results of Sgr~A* by the EHT, its estimated mass is $M\approx 4 \times 10^{6} \; M_{\odot}$~\cite{Event_Horizon_Telescope_Collaboration_2022}. 
As for the mass accretion rate, Faraday rotation measurements of polarized millimeter emission from the vicinity of the event horizon suggest an accretion rate of $10^{-9} - 10^{-7} \; \rm M_{\odot} \, yr^{-1}$ \cite{Marrone_2006}. 
On the other hand, recent EHT observations combined with numerical simulations predict an accretion rate of $\sim 10^{-8} \; \rm M_{\odot} \, yr^{-1}$~\cite{Event_Horizon_Telescope_Collaboration_2022_V}, which is the value we adopt in our model. While the BH mass and accretion rate are estimated from observations, the remaining model parameters need to be specified. 
The viscosity parameter is typically expected to lie in the range $\alpha \sim 0.1 - 0.3$ for ADAF solutions, with values $\sim 0.2-0.3$ commonly adopted in the literature (e.g.,~\cite{Narayan_2008, 1998Narayan}), thus we choose $\alpha=0.3$. 
The $\beta$ parameter, defined as $\beta = P_{g}/(P_{g} + P_{m})$, lies in the range $\sim 0.5 - 0.9$ \cite{Yuan_2014}. We assume $\beta=0.5$, corresponding to exact equipartition between the magnetic field and the gas pressure, consistent with previous works (e.g.,~\cite{1998Narayan, Feng_2017, Oh_2024}). 
Finally, the advection parameter $f_{\rm adv}$ ranges between 0 and 1, with $f_{\rm adv} \to 0$ corresponding to cooling-dominated flows (negligible advection) and $f_{\rm adv} = 1$ to fully advection-dominated flows (negligible radiative cooling). In this work, we take $f_{\rm adv}= 0.9994$, following the results from Ref.~\cite{1998Narayan}.
We consider that the ADAF extends from $\rm \tilde{r}=1$ to $\rm \tilde{r}=10^5$. For Sgr~A*, the outer boundary of the accretion flow is commonly associated with the Bondi radius, $R_{\rm B}\sim10^5 \, R_{\rm S}$, where the thermal energy of the ambient gas becomes comparable to its gravitational binding energy in the potential of the BH~\cite{Yuan_2002}. More generally, the outer boundary of an accretion flow can also be related to the radius at which the outward viscous transport of angular momentum is balanced by external torques. The precise location of this transition in Sgr~A* remains uncertain and is expected to be related to the region from which the gas supply originates~\cite{1998Narayan}. Therefore, $\rm \tilde{r} = 10^5$ provides a physically motivated representative value for the outer extent of the flow.
Nevertheless, since the energetically dominant region lies much closer to the BH, our results are not sensitive to the precise choice of the outer radius. In contrast, local ADAF parameters such as $\alpha$ and $\beta$ affect the density and temperature structure of the flow. Hence, the values adopted here should be regarded as representative of the ranges commonly considered for ADAFs.

\subsection{Magnetically arrested disk (MAD)}
\label{subsec:MAD}
General relativistic magnetohydrodynamics (GRMHD) simulations of ADAF models performed by the EHT collaboration show that the accretion flow can settle into different modes depending on the magnetic field configuration~\cite{Event_Horizon_Telescope_Collaboration_2022, Event_Horizon_Telescope_Collaboration_2022_V}. They distinguish between the standard and normal evolution (SANE;~\cite{Narayan_2012}) and MAD states~\cite{Narayan_2003_MAD, Igumenshchev_2003}. In the SANE scenario, the magnetic field remains relatively weak and turbulent, with magnetic pressure subdominant with respect to the gas pressure, thus not strongly affecting the global dynamics of the flow. In contrast, the MAD state arises when a significant amount of poloidal magnetic flux accumulates near the BH, leading to magnetic pressure comparable to or exceeding the ram pressure of the gas. In this regime, the strong and more ordered magnetic fields can even disrupt the accretion flow, producing intermittent inflow and enabling a high energy extraction efficiency, often associated with the launching of powerful jets.

Recent observational results of Sgr~A*~\cite{Event_Horizon_Telescope_Collaboration_2022} favor the MAD scenario, suggesting a magnetically dominated accretion environment. In particular, the only models that satisfy all EHT constraints — except those related to variability — are MAD models with BH spins of 0.5 and 0.94, both assuming that the BH has an inclination (between the spin axis and the line of sight) of $i=30^{\circ}$. Therefore, we will adopt an ADAF/MAD model to describe the accretion disk of Sgr~A*. 
Besides,  observational data from the EHT collaboration reveals field strengths of $26^{+3}_{-4} \; \rm G$ at $\sim 3.7 \; R_{\rm S}$, $67^{+8}_{-9} \; \rm G$ at $ 2 \; R_{\rm S}$, and $560^{+80}_{-80} \; \rm G$ at the vicinity of the event horizon~\cite{2024ETH_VIII}. 

\begin{table}[h]
    \centering
    \begin{tabular}{ccc}
        \hline
         Model & $a_\ast$ & $B \; \rm [G]$ \\
         \hline 
         \texttt{LOW-B} & 0.94 & 30 \\
         \texttt{Benchmark} & 0.94 & 100 \\
         \texttt{HIGH-B} & 0.94 & 560 \\
         \texttt{LOW-a} & 0.5 & 100 \\
         \texttt{MID-a} & 0.7 & 100 \\
    \end{tabular}
    \caption{Summary of the representative scenarios considered in this work, defined by different combinations of dimensionless BH spin parameter $a_\ast$ and magnetic field strength $B$.}
    \label{tab:scenarios}
\end{table}

In this work, we explore a set of representative scenarios by considering values of the dimensionless BH spin parameter in the interval $a_\ast \in [0.5, 0.94]$ and magnetic field strengths ($B$) of $30 \; \rm G$, $100 \; \rm G$ and $560 \; \rm G$. We define our \texttt{Benchmark} scenario as the case with $a_\ast = 0.94$ and $B = 100 \, \rm G$, which corresponds to an intermediate magnetic field strength and a high spin value favored by recent observations. To assess the impact of the magnetic field, we consider two additional configurations at fixed spin $a_\ast = 0.94$, namely the \texttt{LOW-B} ($B = 30 \; \rm G$) and \texttt{HIGH-B} ($B = 560 \; \rm G$) cases, representing the lower and upper ends of the expected range. In addition, we explore the dependence on the BH spin by fixing the magnetic field to $B = 100 \; \rm G$ and varying the spin parameter, defining the \texttt{LOW-a} ($a_\ast = 0.5$) and \texttt{MID-a} ($a_\ast = 0.7$) scenarios.
A summary of all configurations is provided in Table~\ref{tab:scenarios}.

\section{Neutron production in the accretion disk}
\label{sec:neutron}

For a quantitative estimate of the MPP in Sgr~A*, we consider neutron beta decay occurring within its ergosphere: 
\begin{equation}
   n \rightarrow p + e^{-} + \bar{\nu}_{e}.
    \label{eq:beta_decay}
\end{equation}

In order to determine the population of neutrons in the ergosphere capable of undergoing the MPP, it is first necessary to evaluate the neutron production rate within the accretion flow. In the ADAF framework, the plasma reaches sufficiently high ion temperatures to enable neutron production through nuclear reactions. In this Section, we compute the corresponding neutron production rate in the accretion flow of Sgr~A*, accounting for the gravitational escape, which reduces the number of neutrons retained in the flow and thus affects the population available to reach the ergosphere.

\subsection{Nuclear reactions}
\label{subsec:nuclear_react}

In the ADAF scenario, ions — primarily protons and $\alpha$ particles (i.e., free $^4$He nuclei) — can reach approximately virial temperatures, rising up to $\sim 10^{12} \;\rm K$ in the innermost regions of the flow, whereas electrons remain significantly cooler, with temperatures in the range $10^{9}$–$10^{11} \, \rm K$.\footnote{Electrons are subject to efficient radiative cooling processes, such as bremsstrahlung, synchrotron emission, and inverse Compton scattering~\cite{Narayan_1995b}. Consequently, their temperature remains comparatively low, suppressing reactions such as electron capture and justifying its neglection in our study.} As a result, the plasma is ion-dominated, and nuclear interactions involving protons and $\alpha$ particles become efficient. 

\begin{widetext}

\begin{figure}[h]
\begin{center}
		\begin{subfigure}{0.49\textwidth}
			\includegraphics[width=\textwidth]{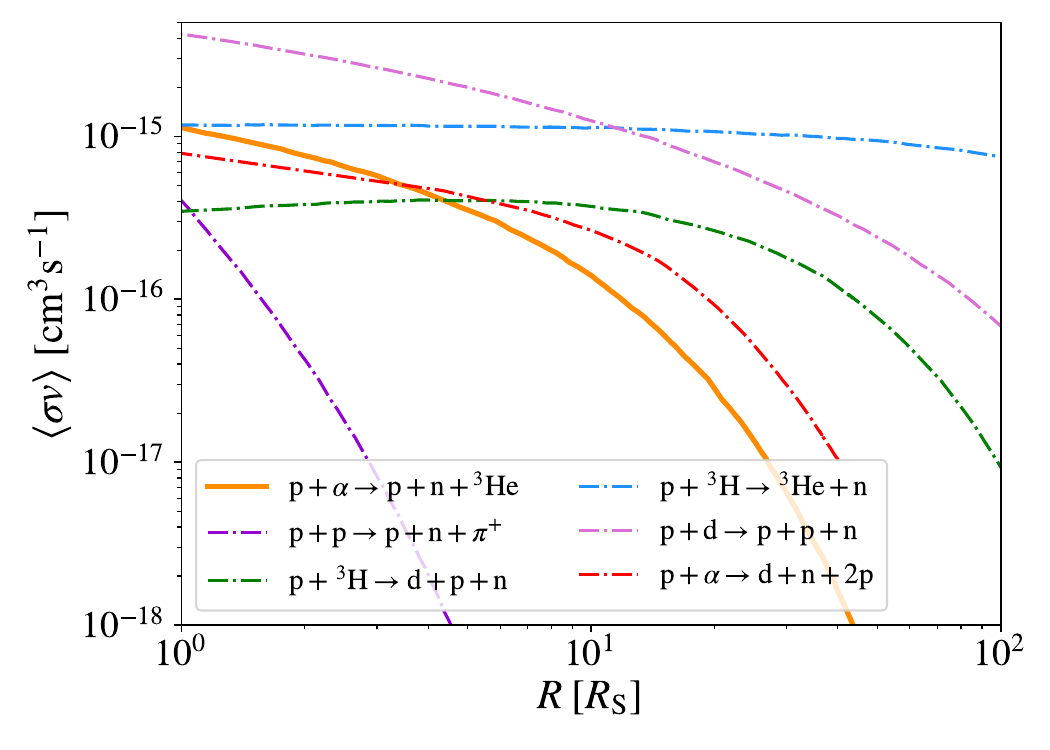}
			
		\end{subfigure}
		\hspace{0.01cm}
		  \begin{subfigure}{0.49\textwidth}
		      	\includegraphics[width=\textwidth]{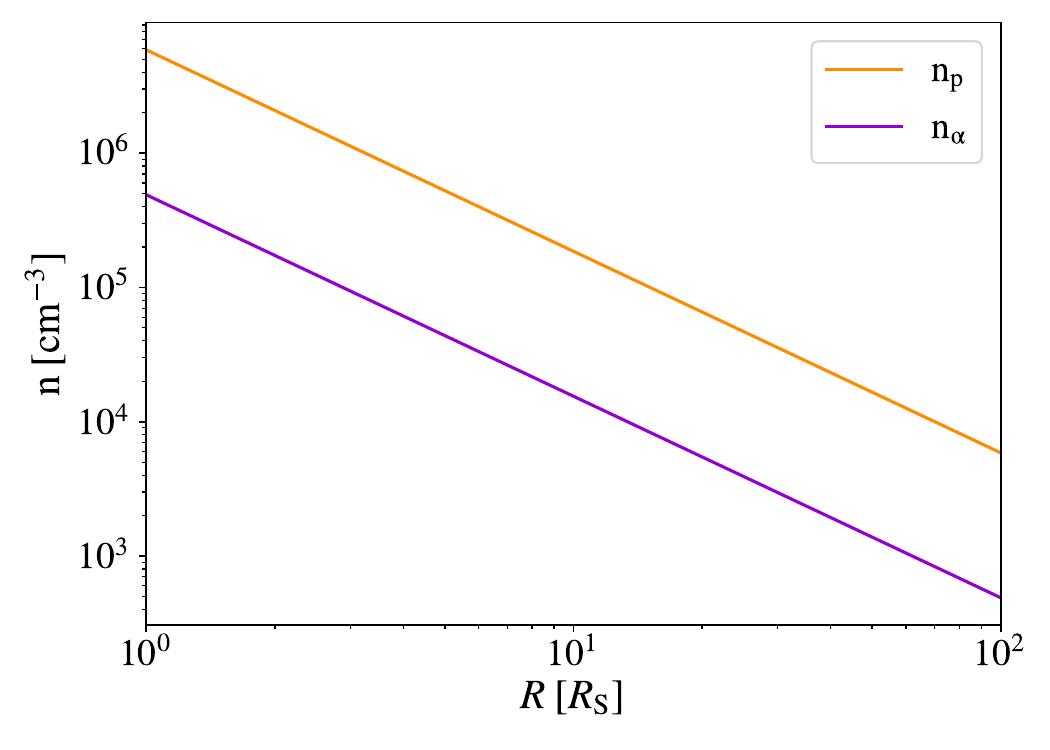}
			
		  \end{subfigure}
        \caption{Left panel: Velocity-averaged cross-section values, $\langle \sigma v \rangle$, for the main nuclear reactions producing neutrons in an ADAF as a function of the disk radius (see Section~\ref{subsec:nuclear_react} for details.). Right panel: Radial profiles of the proton ($\rm n_p$) and $\alpha$-particle ($\rm n_\alpha$) number densities in the accretion flow of Sgr~A*, computed within the ADAF framework using the mass density profile given by Eq. \eqref{eq:rho}, and assuming hydrogen and helium mass fractions $X_{\rm H}=0.75$ and $X_{\rm He}=0.25$, respectively. We adopt $\alpha=0.3$, $\beta=0.5$, and $f_{\rm adv}=0.994$. }
        \label{fig:nuclear_reactions}
		
\end{center}
\end{figure}
\end{widetext}

In the left panel of Figure~\ref{fig:nuclear_reactions}, we display the velocity-averaged cross-sections for the main network of nuclear reactions expected to occur in an ADAF (see, e.g., Refs.~\cite{1990Guessoum, Guessoum_1999} for details). We adopt the cross-sections for the reactions $p+\alpha \rightarrow p+n+{}^{3}\mathrm{He}$, $p+p \rightarrow p+n+\pi^{+}$, $p+{}^{3}\mathrm{H} \rightarrow d+p+n$, $p+{}^{3}\mathrm{H} \rightarrow {}^{3}\mathrm{He}+n$, and $p+d \rightarrow p+p+n$ from Ref.~\cite{1990Guessoum}, while the cross-section for $p+\alpha \rightarrow d+n+2p$ is taken from Ref.~\cite{Aharonian1984}.

Since the velocity-averaged cross-sections are of comparable magnitude in the innermost region, the threshold energy required for each reaction becomes the key factor determining their relative importance. Moreover, assuming a reasonable composition for the accretion plasma, with hydrogen and helium mass fractions of $X_{\rm H} = 0.75$ and $X_{\rm He} = 0.25$, respectively~\cite{Narayan_1995b}, and negligible abundances of other species such as $^{3}\mathrm{H}$ or deuteron ($d$), neutron production is expected to be dominated by reactions involving protons and $\alpha$ particles. The right panel of Figure~\ref{fig:nuclear_reactions} shows the number density of protons and $\alpha$ particles as a function of the radius, illustrating that the density of both species increases significantly toward smaller radii, thus enhancing the efficiency of neutron production in the inner regions. In particular, neutrons are primarily generated through $\alpha$-particle dissociation, with an additional contribution from the $p+p \rightarrow p+n+\pi^{+}$ channel at sufficiently high ion temperatures.
Note, also, that given the high temperatures required for these reactions to occur, this mechanism is expected to operate mainly in the inner regions of the flow, at radii $\lesssim 100 \; R_{\rm S}$~\cite{1990Guessoum}. Therefore, in the following we compute the neutron production rate focusing on the dominant channel, $p+\alpha \rightarrow p+n+{}^{3}\mathrm{He}$, as the inclusion of subdominant reactions does not significantly affect our results.

\subsection{Neutron production rate}
\label{subsec:neutron_rate}

Previous studies aimed at estimating the neutron production rate in ADAFs (e.g.,~\cite{Guessoum_1999, Jean_2001, Oh_2024}) typically assume that neutrons follow an MB distribution. 
This assumption relies on the expectation that Coulomb interactions efficiently thermalize the ions on timescales much shorter than those associated with nuclear reactions, leading to approximately Maxwellian particle distributions. 
In this context, in Ref.~\cite{Guessoum_1999} it is argued that adopting a thermal distribution for neutrons, with a temperature equal to that of the ions, provides a reasonable approximation. 
However, while this assumption is well justified for protons and $\alpha$ particles — whose distributions are continuously driven toward thermal equilibrium by frequent Coulomb interactions — the same does not necessarily hold for neutrons. 
Although neutrons can in principle exchange energy through nuclear interactions, they lack the efficient Coulomb coupling that rapidly thermalizes charged particles. Therefore, the validity of an MB approximation for neutrons depends on whether such interactions can efficiently redistribute their energy before escaping or decaying. 
Due to the low baryonic density of ADAFs, the produced neutrons are effectively collisionless. Using the local proton density and a representative neutron–proton scattering cross-section~\cite{1990Guessoum}, we find that the neutron mean free path exceeds the assumed radial extent of the flow ($\lambda_n > 10^{5} \, R_{\rm S}$).
The corresponding scattering optical depth is therefore smaller than unity, implying that most neutrons undergo no further collisions after production and cannot establish thermal equilibrium with the ambient ions.

In our approach, protons and $\alpha$ particles are likewise assumed to be thermalized. 
However, the large neutron mean free path suggests that thermalization of neutrons is inefficient. Consequently, we do not assume an MB distribution for neutrons and instead compute their spectrum directly from the nuclear production process of the relevant reaction ($p+\alpha \rightarrow p+n+{}^{3}\mathrm{He}$).
Rather than relying on thermally averaged reaction rates, this treatment allows us to determine the energy distribution of the produced neutrons from the underlying interaction kinematics, providing a more accurate description of the neutron population in the accretion flow. 

We then evaluate the neutron production rate in the accretion flow of Sgr~A*. Since neutron production is most efficient in the innermost regions of the accretion flow (see Figure~\ref{fig:nuclear_reactions}), we restrict our study to $R \le 15 \, R_{\rm S}$. 
The neutron production rate, i.e., the number density of neutrons produced per unit energy and time, is obtained from the collision term of the Boltzmann equation (see, e.g., Ref.~\cite{RAFFELT19901}):

\begin{widetext}
\begin{equation}
    \frac{d^2 n_n}{dE_n dt}  =  \prod_i \int \frac{g_id^3 \vec{p}_i}{(2 \pi)^3 2 E_i} f_i (E_i)   \prod_{j \neq n} \int \frac{d^3 \vec{p'}_j}{(2 \pi)^3 2 E'_j} [1 - f_j (E'_j) ] \, (2 \pi )^4 \delta^4 \Bigl( \sum_k p_k - \sum_{l \neq n} p'_l-p_n \Bigr) \frac{|{\vec{p}_n}|}{4 \pi^2} |{\mathcal{\overline{M}}|}^2.    
    \label{eq:prod_spectrum}
\end{equation}
\end{widetext}
 
\noindent Here $p_i=(E_i,\vec{p}_i)$ are the four-momenta of the incoming particles, 
where $E_i$ and $\vec{p}_i$ denote their energies and three-momenta, respectively, and $g_i$ their degrees of freedom. The four-momenta of the outgoing particles, excluding the neutron, are labeled as $p'_j=(E'_j,\vec{p'}_j)$, 
while $p_n=(E_n,\vec{p}_n)$ refers specifically to the produced neutron. Finally, $|\overline{\mathcal{M}}|^2$ is the spin-averaged squared matrix element for the neutron-producing processes. The functions $f_i (E_i)$ and $f_j (E'_j)$ are the distribution functions for the incoming and outgoing particles, respectively, which follow a Fermi-Dirac distribution for protons, neutrons and ${}^{3}\mathrm{He}$,  and a Bose-Einstein distribution for the $\alpha$ particles.

The expression of Eq.~\eqref{eq:prod_spectrum} can be rewritten as:

\begin{widetext}
\begin{equation}
\frac{d^2 n_n}{dE_n dt}  =     \int 2 \frac{ d^3 \vec{p}_p}{(2 \pi)^3} f_p (E_p) \int \frac{d^3 \vec{p}_\alpha}{(2 \pi)^3 } f_\alpha (E_\alpha) \frac{\sqrt{(p_p p_\alpha)^2-(m_p m_\alpha)^2}}{E_p E_\alpha}  \frac{d\sigma}{dE_n}\left( E_p, E_\alpha, \textrm{cos}\, \theta_{p\alpha}\right),
\label{eq:neutron_prduction_rate}
\end{equation}
\end{widetext}

\noindent where $\frac{d\sigma}{dE_n}\left( E_p, E_\alpha, \textrm{cos}\, \theta_{p\alpha}\right)$ is the differential cross-section for the process producing neutrons, with $\theta_{p\alpha}$ the angle between the three-momenta of the incoming proton and the $\alpha$-particle. This differential cross-section is obtained by transforming the laboratory-frame expression of Ref.~\cite{Jung1973} to a general reference frame (see Appendix~\ref{app:cross_section_diff}). Note that we have neglected Pauli blocking for the outgoing particles, i.e., we set $[1-f_j(E'_j)] \approx 1$, as we assume that the plasma is non-degenerate.

\begin{figure}[h!]
    \centering
    \includegraphics[width=1.0
    \linewidth]{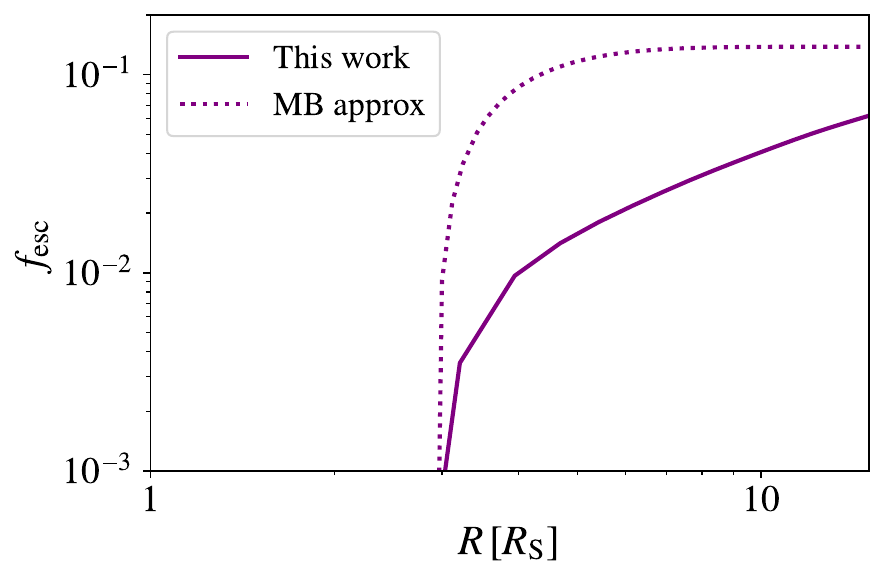}
    \caption{Escape fraction of neutrons, defined as the fraction of neutrons with kinetic energies exceeding the gravitational binding energy, shown as a function of distance from the BH. The solid line shows the result obtained from the neutron energy distribution computed in this work, while the dotted line shows the corresponding estimate under the MB approximation.}
    \label{fig:f_esc}
\end{figure}

Once produced, neutrons can escape from the accretion flow provided their energies exceed the local gravitational binding energy. In previous works (e.g., ~\cite{Aharonian1984, 1990Guessoum, Oh_2024, Jean_2001}), the escape fraction is estimated by counting the neutrons with sufficient energy to overcome the gravitational potential, under the assumption that their velocities follow a thermal Maxwellian distribution. In contrast, here we compute the escape fraction using the energy distribution derived above, by directly evaluating the fraction of neutrons whose kinetic energy exceeds the gravitational binding energy, and show the results in Figure~\ref{fig:f_esc}. 
For comparison, the same quantity is also shown assuming an MB distribution for the neutron energies.
As expected, neutron losses are negligible at small radii, where the gravitational potential is deepest. However, as the production radius increases, the escape fraction becomes non-negligible, indicating that an increasing fraction of neutrons can overcome the gravitational potential. In particular, at $R \sim 15 \, R_{\rm S}$ we find $f_{\rm esc} \sim 6\times10^{-2}$, implying that only about $6\%$ of the produced neutrons escape, while the vast majority remain gravitationally bound. This effect is expected to become more pronounced at larger radii. 
The MB approximation yields systematically larger escape fractions than those obtained from the reaction kinematics, by a factor of a few over most of the radial range considered.
This difference arises from retaining the neutron energy distribution set by the production process, rather than assuming thermal equilibrium with the surrounding plasma. Consequently, the thermal approximation may substantially overestimate the number of neutrons able to escape the gravitational potential of the accretion flow.

\begin{figure}[h!]
    \centering
    \includegraphics[width=1.0
    \linewidth]{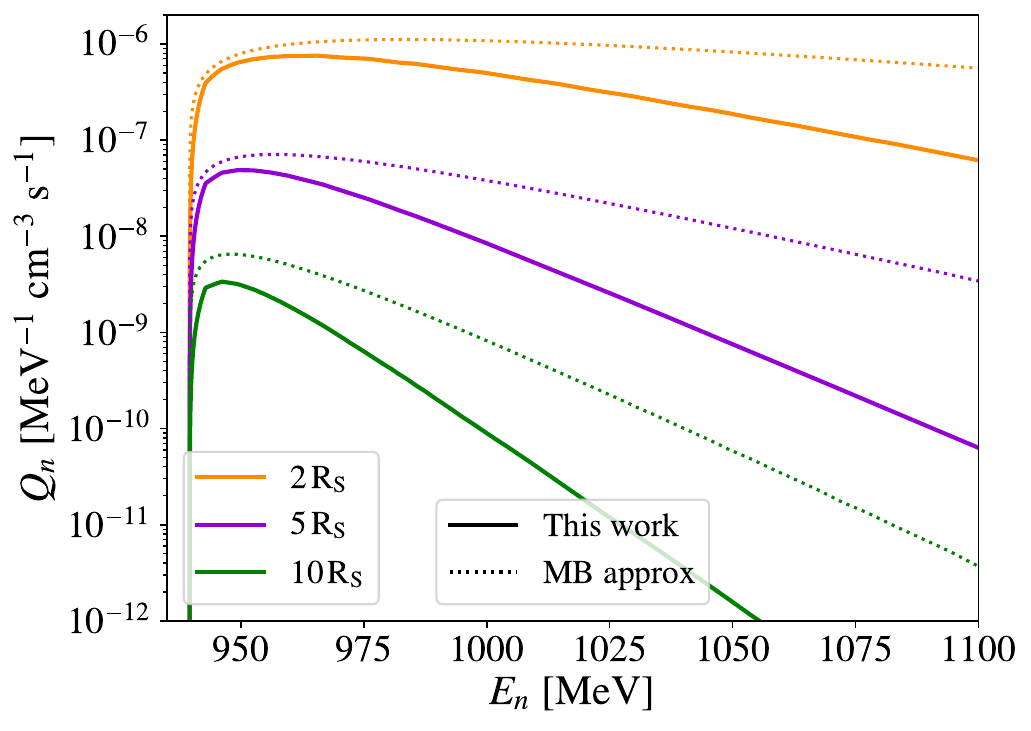}
    \caption{Neutron production spectrum rate in the accretion flow of Sgr~A* as a function of the neutron energy, shown for different production radii and corrected for losses due to gravitational escape. Solid lines correspond to the spectra obtained in this work by explicitly accounting for the neutron energy distribution from the underlying interaction kinematics, while dotted lines show the spectra obtained under the standard MB approximation (see Section~\ref{subsec:neutron_rate} for full details).}
    \label{fig:inicial_neutron_rate}
\end{figure}

After accounting for the gravitational escape, we obtain the differential neutron production rate in the accretion flow of Sgr~A* as:
\begin{equation}
    Q_{n} = \frac{ d^2 n_n}{ dE_n \,dt} \times (1-f_{\mathrm{esc}}).
    \label{eq:Q_n}
\end{equation}
Figure~\ref{fig:inicial_neutron_rate} shows in solid lines the neutron production spectrum at different radii within the accretion flow after accounting for gravitational escape, and preserving the energy dependence imposed by the reaction kinematics. It illustrates that neutron production is strongly dominated by the innermost regions of the accretion flow, where the higher plasma densities and enhanced reaction efficiencies lead to a significantly larger neutron production. For comparison, we also include the spectra obtained under the conventional MB approximation in dotted lines. While both approaches predict similar overall spectral shapes, noticeable differences appear in the normalization and energy dependence, especially at larger radii, reflecting the impact of explicitly accounting for the reaction kinematics rather than relying on thermally averaged reaction rates. 

\section{Magnetic Penrose Process (MPP) in Sgr A*}
\label{sec:MPP}
In this Section, we present a brief theoretical description of the MPP, a physical mechanism that can operate in the ergosphere of rotating 
BHs as a result of the interaction between charged particles and the magnetic fields generated by the surrounding plasma. We also introduce the Wald solution \cite{PhysRevD.10.1680} as an approximate description of the magnetic field configuration in Sgr~A*.
Building on this framework, we estimate the population of neutrons reaching the ergosphere, being those able to undergo MPP.
Finally, we compute the efficiency of the MPP for the different scenarios considered in this work, and derive the resulting spectrum of protons accelerated via the MPP in Sgr~A*.

\begin{figure}[h!]
\includegraphics[width=0.9
\linewidth]{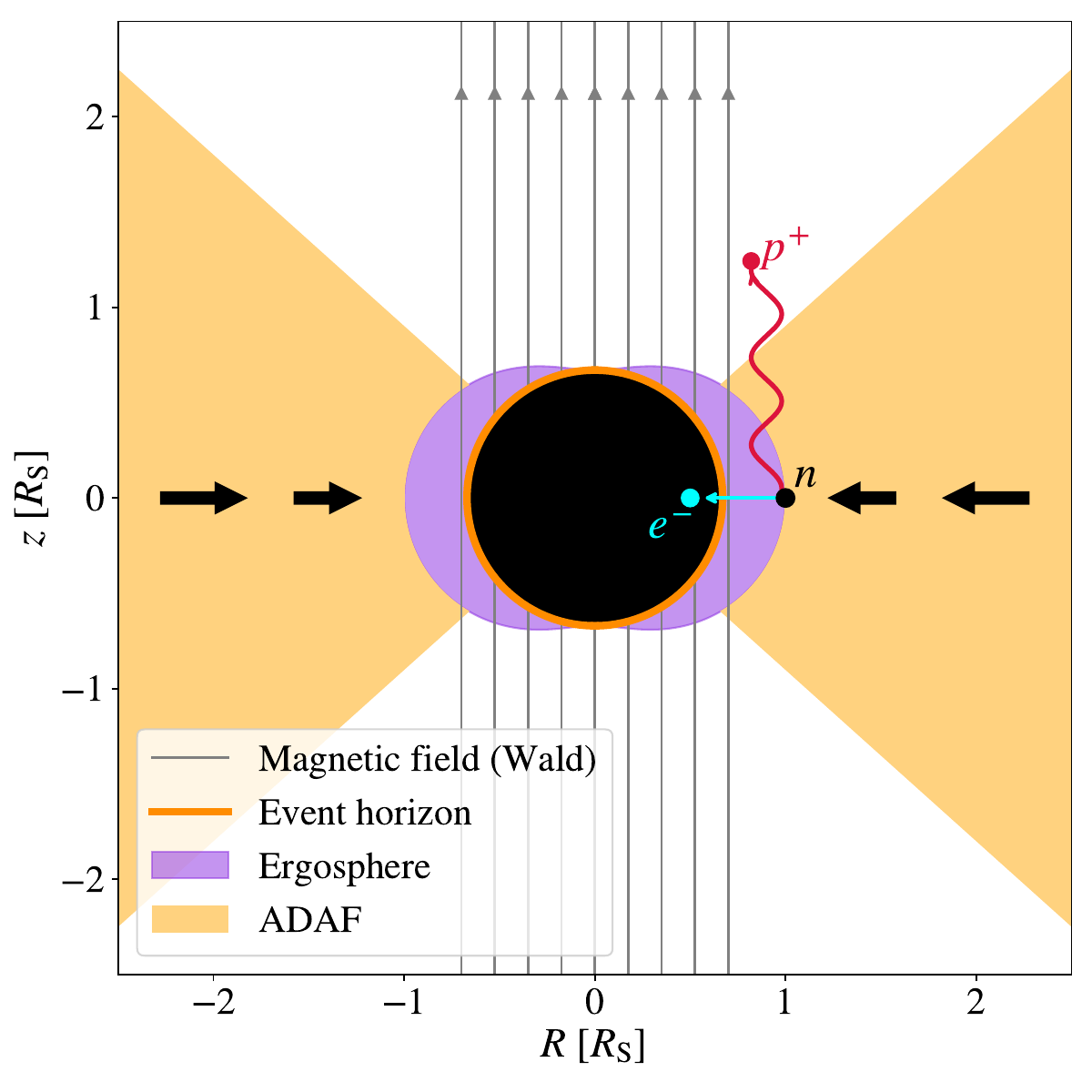}
\caption{Schematic representation of a system composed of a rotating BH (in this case, Sgr A$^*$) surrounded by an accretion flow described by the ADAF model (orange), which extends inward down to the ergosphere (purple). The figure illustrates the model considered here: neutrons (black point) produced in the accretion flow penetrate the ergosphere (as the black arrows show), where they undergo neutron beta decay into a proton (red), an electron (cyan), and an antineutrino (not shown). Within the ergosphere, energy extraction becomes possible: if one of the decay products (the electron, in this example) falls into the event horizon (dark orange), the other particle can escape to infinity with increased energy, consistent with a Penrose-type process. In the presence of an external magnetic field described here by the Wald solution (gray field lines), the charged particles interact with the electromagnetic field, enhancing the efficiency of the process. This mechanism is commonly known as the MPP. Note that the trajectory of the escaping proton is along the magnetic field lines, which are collimated along the polar direction ($z$).}
\label{fig:structure}
\end{figure}

Our final setup is illustrated in Figure~\ref{fig:structure} and can be summarized as follows. Neutrons are first produced within the accretion flow of Sgr~A*, which is well-described by an ADAF model. We then compute the subset of neutrons that is able to reach the ergosphere (Section~\ref{subsec:neutrons_ergo}). Once inside this energy extraction region, neutrons undergo beta decay, producing a proton, an electron, and an antineutrino. We neglect the contribution of the antineutrino and, under the magnetic field configuration described by the Wald solution (Section~\ref{subsec:wald}), we assume that electrons are captured by the BH, while protons escape to infinity with energies exceeding that of the parent neutron. 

\subsection{Theoretical framework}
\label{subsec:theoretical_fram}

\subsubsection{MPP in a nutshell}
\label{subsec:MPP_theory}
A rotating BH is described by the Kerr geometry, which in Boyer–Lindquist coordinates
takes the form \cite{Boyer1967}:

\begin{equation}
\begin{split}
ds^2 = & -\left(1 - \frac{2 M r}{\Sigma}\right) dt^2
- \frac{4 M a r \sin^2\theta}{\Sigma} \, dt \, d\phi
+ \frac{\Sigma}{\Delta_r} \, dr^2 \\
& + \Sigma \, d\theta^2 + \left(r^2 + a^2 + \frac{2 M a^2 r \sin^2\theta}{\Sigma}\right)
  \sin^2\theta \, d\phi^2,
\end{split}
\label{metric}
\end{equation}
where:
\begin{equation}
    \Sigma = r^2 + a^2 \cos^2\theta, \qquad
    \Delta_r = r^2 - 2 M r + a^2.
    \label{eq:Delta_kerr}
\end{equation}

The spacetime is fully determined by the BH mass $M$ and angular momentum $J$, both related through the spin parameter $a=J/M$, with $0\le a \le M$. We also define the dimensionless spin parameter as $a_\ast = a/M$, which satisfies $0\leq a_\ast\leq1$. The Kerr geometry possesses a ring singularity located at $\Sigma=0$, corresponding to $r=0$ and $\theta=\pi/2$.

A defining feature of rotating BHs is the presence of event horizons, determined by the condition $g^{rr}=0$, or equivalently $\Delta_r=0$. Solving this equation gives:

\begin{equation}
r_{\pm}=M\pm\sqrt{M^2-a^2},
\label{eh}
\end{equation}
which correspond to the inner ($r_-$) and outer ($r_+$) horizons.\footnote{Unlike the Kerr solution, the Schwarzschild BH has a single horizon at $r=2M$.}

The Kerr spacetime also contains the static limit surface, defined by $g_{tt}=0$, where the timelike Killing vector $\xi^\mu_{(t)}$ becomes null. Its location is:

\begin{equation}
r_E(\theta)=M+\sqrt{M^2-a^2\cos^2\theta}.
\label{eqergo}
\end{equation}

In the equatorial plane, the static limit remains fixed at $r_E(\theta=\pi/2)=2M$, which corresponds to the Schwarzschild radius $R_{\rm S}$. The region between the static limit and the outer event horizon defines the ergosphere, where no observer can remain stationary due to frame dragging, and all physical trajectories are forced to co-rotate with the BH.
 
The existence of an ergosphere is a necessary condition for any Penrose-type energy extraction process. To illustrate this, consider a particle approaching a rotating BH from infinity. The spacetime admits a timelike Killing vector field $\xi^\mu$, associated with time translations as measured by a distant observer. Outside the ergosphere, this vector is timelike, satisfying $\xi^\mu \xi_\mu < 0$. However, within the ergoregion, $\xi^\mu$ becomes spacelike, such that $\xi^\mu \xi_\mu > 0$.

For a particle with four-momentum $p^{\mu}$ following a geodesic, the quantity

\begin{equation}
p^\mu \nabla_\mu (\xi^\nu p_\nu) = 0
\end{equation}
vanishes as a consequence of both the Killing equation and the geodesic equation \cite{1997gr.qc....12019C}. This implies the conservation along the trajectory of the quantity

\begin{equation}
E = - p_\mu \xi^\mu,
\end{equation}
which is interpreted as the energy of the particle as measured by an observer at infinity.
 
Outside the ergosphere, where $\xi^\mu$ is timelike, the conserved energy is positive definite for future-directed timelike trajectories. Nevertheless, inside the ergoregion, where $\xi^\mu$ becomes spacelike as a consequence of the frame dragging, the quantity 
$E$ is no longer constrained to be positive. As a result, negative-energy states (as measured at infinity) can exist within the ergosphere, providing the essential condition for the Penrose-type energy extraction processes. This possibility was first recognized by Roger Penrose in 1969 \cite{Penrose:1969pc}, who showed that particles within the ergoregion may occupy negative-energy orbits with respect to an observer at infinity, thereby enabling the extraction of rotational energy from the BH. 

Consider a particle entering the ergosphere from infinity with positive energy $(E_1 > 0)$. Inside this region, the particle decays into two fragments, and energy conservation implies:
\begin{equation}
    E_1 = E_2 + E_3.
\end{equation}

As discussed above, the conserved energy within the ergosphere is not positive definite. Consequently, one of the fragments can follow a trajectory with negative energy ($E_2 < 0$), while the other retains positive energy ($E_3 > 0$). If the negative-energy particle is captured by the BH and crosses the event horizon, conservation of energy requires that the escaping fragment carries away more energy than the original particle, i.e.,
\begin{equation}
    E_3 > E_1.
\end{equation}
This excess energy is extracted from the rotational energy of the BH \cite{1972ApJ...178..347B}. Therefore, the existence of negative-energy states within the ergosphere provides the fundamental mechanism underlying the PP.

However, the original PP is characterized by a very low efficiency and requires unrealistic kinematic conditions. In realistic astrophysical environments, electromagnetic fields play a crucial role and can significantly enhance the efficiency of energy extraction. This naturally leads to the MPP~\cite{1985ApJ...290...12W,1986ApJ...307...38P,1985JApA....6...85B,1989PhR...183..137W}. 
In the presence of an external magnetic field, charged particles no longer follow purely geodesic motion, but instead experience Lorentz forces caused by the magnetic field. Consequently, their conserved quantities are modified. In particular, the conserved energy associated with $\xi^\mu$ becomes:

\begin{equation}
E = - p_\mu \xi^\mu - q A_\mu \xi^\mu,
\end{equation}
where $q$ is the particle charge and 
$A_\mu$ is the electromagnetic four-potential. This additional interaction term allows for a wider range of negative-energy states within the ergosphere. As a result, the energy extraction mechanism becomes significantly more efficient than in the purely gravitational PP.

Several studies suggest that electromagnetic interactions effectively extend the energy extraction region beyond the ergosphere, giving rise to what can be interpreted as an \textit{effective} ergosphere \cite{Dhurandhar1984_I, Dhurandhar1984_II, Tursunov_2019}. While the classical ergosphere is geometrically bounded at $R_{\rm S}$ in the equatorial plane, the inclusion of electromagnetic terms in the effective potential of the particles (particularly the $-qA_t$ contribution) allows for the existence of negative-energy orbits beyond this limit. Since such orbits are the key requirement for energy extraction mechanisms like the MPP, this extended region significantly enhances both the efficiency and spatial domain of the process. Nevertheless, as discussed later in Section~\ref{subsec:neutrons_ergo}, extending the energy extraction region has only a minor impact on our final results. We therefore restrict our analysis to particle decays occurring within the classical ergosphere.

\subsubsection{Magnetic field configuration}
\label{subsec:wald}

BHs are commonly surrounded by ionized accretion disks, where electric currents in the plasma generate magnetic fields. In binary systems, BHs can be immersed in the magnetic field of a companion star, which can be particularly strong if the companion is a neutron star or magnetar. Even isolated BHs can be immersed in weak large-scale magnetic fields of galactic or extragalactic origin.

As introduced in Section~\ref{subsec:MAD}, Sgr~A* is believed to reside in a strongly magnetized accretion environment. In this work, we assume that the accretion flow of Sgr~A* is in a MAD state, in which magnetic flux accumulates near the BH horizon, leading to a highly magnetized region. Although the magnetic field configuration in this region is expected to be complex, we consider that, within the small spatial region where particle decays occur, the field can be approximated as locally uniform. Under this assumption, the Wald solution \cite{PhysRevD.10.1680} provides a useful analytical description of the electromagnetic field.
In particular, the Wald solution corresponds to an exact solution of the Maxwell's equations in a Kerr spacetime, describing a uniform magnetic field aligned with the rotation axis of the BH.

The associated electromagnetic four-potential of the
electromagnetic field $A_{\mu}$ has two non-vanishing covariant components $A_t$ and $A_{\phi}$ that can be written as \cite{PhysRevD.10.1680}:

\begin{equation}
    A_t=aB\left(\frac{Mr}{\Sigma}(1+\cos ^2\theta)-1\right),
\label{waldpotentialt}
\end{equation}

\begin{equation}
    A_\phi=\frac{B}{2}\left(r^2 + a^2 + \frac{2 M a^2 r}{\Sigma} (1+\cos ^2\theta)\right)
  \sin^2\theta.
\label{waldpotentialphi}
\end{equation}

The BH rotation generates the quadrupole electric field given by $A_t$, arising from the twisting of magnetic field lines due to frame dragging. This electromagnetic contribution provides the necessary energy for particles to access negative-energy orbits, removing the stringent constraints on the relative velocities between fragments required in the original PP.

The induced charge will correspond to the induced BH charge. As shown in Ref.~\cite{Tursunov_2020}, the presence of an induced BH charge plays a crucial role in the dynamics of charged particles. In its absence, particles tend to be captured by the BH. By contrast, when an induced charge is present, the electromagnetic interaction can supply the required negative energy to one of the fragments, thus allowing the other to escape to infinity with enhanced energy.

In realistic astrophysical environments, the most relevant configuration is the one in which the magnetic field is aligned with the BH spin axis, at least in its vicinity, leading to a positive temporal induced charge (e.g., \cite{Zaja_ek_2018, zajacek2019}). In our scenario, which involves neutron beta decay, the escaping particle could in principle be either the electron or the proton. However, under the assumption of a positively charged BH, the energetically favored outcome is that electrons are captured by the BH, while protons escape to infinity with enhanced energy. Since charged particles are expected to follow the magnetic field lines, the escaping protons will preferentially propagate along the rotation axis, i.e., in the polar direction. We therefore adopt this configuration in the following. 

\subsection{Neutrons reaching the ergosphere}
\label{subsec:neutrons_ergo}

In Section~\ref{sec:neutron}, we derived the neutron population within the accretion flow of Sgr~A*. However, our focus here is on the subset of neutrons that reach the ergosphere, as these are the ones that can ultimately undergo the MPP. Once produced within the accretion flow, we assume that neutrons propagate toward the BH, largely decoupled from the plasma. Being electrically neutral, they are not influenced by electromagnetic fields as in the case of charged particles, and their motion is therefore governed mainly by the gravitational potential and their initial kinematic conditions at production. In addition, as discussed above, the low baryonic densities characteristic of ADAF flows imply that the neutron mean free path greatly exceeds the characteristic scales of the accretion flow, allowing neutrons to be treated as effectively collisionless on the relevant timescales.
As a result, their trajectories can be well approximated as ballistic (i.e., geodesic), determined by the spacetime geometry rather than by the bulk motion of the accretion flow.
 
Under these assumptions, we introduce a self-consistent framework to determine the neutron population reaching the ergosphere, simultaneously accounting for their orbital dynamics in the Kerr spacetime and their finite lifetime due to beta decay.
The orbital dynamics controls whether a neutron can physically reach the target radius, as its trajectory is governed by the spacetime geometry and its conserved quantities, namely the energy and angular momentum. Depending on these parameters, some neutrons follow plunging trajectories that reach the inner regions, while others may encounter turning points or remain on non-infalling orbits, preventing them from accessing the ergosphere. In parallel, even for neutrons whose trajectories allow them to reach the desired radius, their finite lifetime introduces an additional suppression. Since beta decay is governed by the neutron proper lifetime, the relevant timescale is the proper time accumulated along the trajectory rather than the coordinate time. As a result, neutrons that require longer proper times to travel from their production radius to the ergosphere are exponentially less likely to survive. Therefore, the final population of neutrons reaching the ergosphere is determined by the interplay between the accessibility of the trajectory, set by the orbital dynamics, and the survival probability, controlled by the proper time elapsed along it.

For a BH described by the Kerr metric, given by Eq.~\eqref{metric}, the evolution of a particle of mass $m_0$ in the equatorial plane ($\theta=\pi/2$) can be described in terms of its proper time, $\tau$. In particular, the radial component of the geodesic equation can be written as (with $M=1$)~\cite{Bardeen1972, Ba_ados_2009}: 
\begin{equation}
    \frac{dr}{d\tau}= \pm \frac{1}{r^2} \sqrt{T^2 -\Delta (m_0^2 r^2+(l-a_\ast E)^2)},
    \label{eq:proptime}
\end{equation}
where the sign determines the direction of radial motion, $\Delta$ is given in 
Eq.~\eqref{eq:Delta_kerr}, and we have defined $T\equiv E(r^{2}+a_\ast^{2})-la_\ast$. 
Here, $E$ is the total energy of the particle and $l=p_{\phi}=p\sin\theta\sin\phi$ 
is the component of angular momentum parallel to the symmetry axis per unit mass. This expression describes the radial evolution of the particle along timelike geodesics and allows us to compute the proper time elapsed between a certain production radius and the ergosphere. 

The proper time accumulated along the trajectory from an initial radius $r_i$ to a final radius $r_f$ can then be obtained by integrating Eq.~\eqref{eq:proptime}, which provides the characteristic timescale that governs neutron decay along the trajectory. In our case, $r_f = R_{\rm S}$, corresponding to the location of the 
ergosphere in the equatorial plane of Sgr~A*. Since beta decay is governed by the neutron proper lifetime, $\tau_{n}=879.4 \pm 0.6 \, \rm s$~\cite{ParticleDataGroup:2024cfk}, the probability for a neutron to survive without decaying along its trajectory is given by:
\begin{equation}
P_{\rm surv} = \exp\left(-\frac{\tau(r_i \rightarrow r_f)}{\tau_n}\right).
\label{eq:p_surv}
\end{equation}

\begin{figure}[h!]
    \centering
    \includegraphics[width=1.0
    \linewidth]{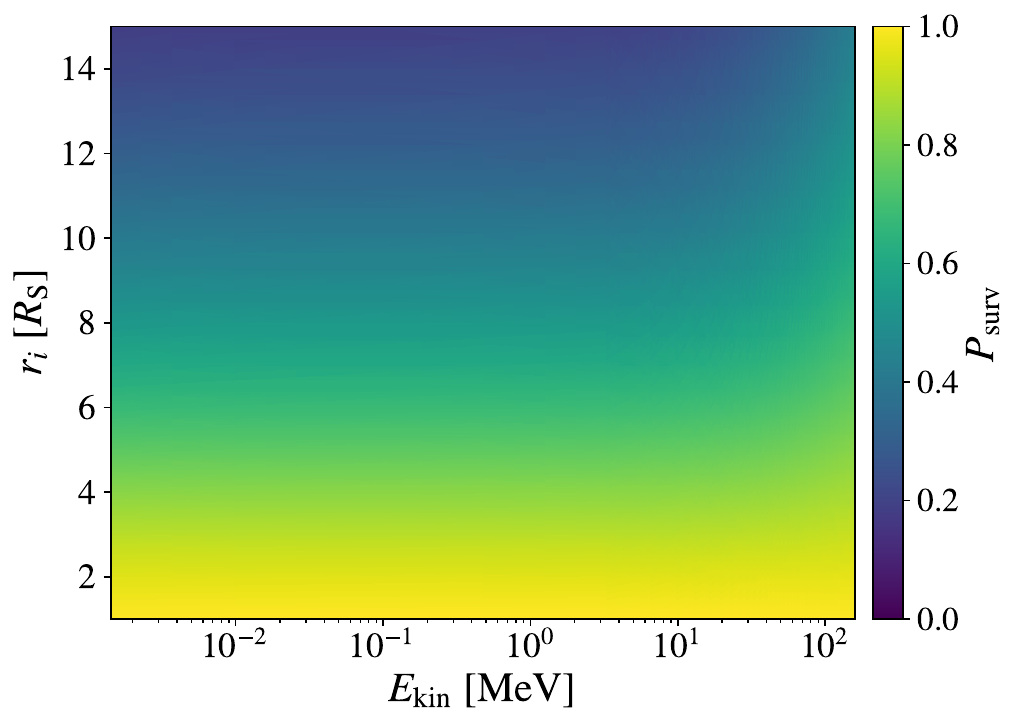}
    \caption{Neutron survival probability, $P_{\rm surv}$, as a function of the kinetic energy, $E_{\rm kin}$, and the initial production radius, $r_i$ (in units of $R_{\rm S}$). The color scale indicates the probability that neutrons produced in the accretion flow of Sgr~A* reach the ergosphere before decaying along their Kerr trajectories. These results correspond to a BH spin of $a_\ast=0.94$.}
    \label{fig:Psurv_kerr}
\end{figure}

Since we adopt a formalism for the calculation of the neutron production rate based on the neutron production cross-section integrated over the emission angles, the resulting neutron spectrum does not explicitly track the direction of the outgoing neutrons. 
Therefore, we assume that neutron emission is isotropic in the local rest frame of the flow, consistent with isotropic parent proton and $\alpha$-particle distributions, and estimate the effect of angular momentum by performing multiple realizations with randomly sampled directions. 
For each neutron energy and production radius, we generate a distribution of $l$ values consistent with isotropic emission in the local frame and compute the corresponding Kerr trajectories. 
For each realization, we evaluate the neutron survival probability based on its proper travel time along the trajectory.
The final survival probability is then obtained by averaging over these realizations, thus incorporating the effect of angular momentum in a statistical way. 
Figure~\ref{fig:Psurv_kerr} illustrates the neutron survival probability as a function of both the kinetic energy and the initial production radius. 
As expected, $P_{\rm surv}$ strongly depends on the production radius. Neutrons produced closer to the ergosphere exhibit higher survival probabilities, as they require shorter travel times to reach this region. In contrast, neutrons originating at larger radii must traverse longer distances, leading to a larger accumulated proper time and, consequently, a stronger suppression due to beta decay. 
The dependence on kinetic energy is comparatively mild. A slight increase of $P_{\rm surv}$ with energy is visible mainly for neutrons produced farther from the ergosphere, since more energetic neutrons have higher velocities and thus accumulate less proper time along their trajectories. As a result, low-energy neutrons produced at large radii are the most strongly suppressed.

\begin{figure}[h!]
    \centering
    \includegraphics[width=1.0
    \linewidth]{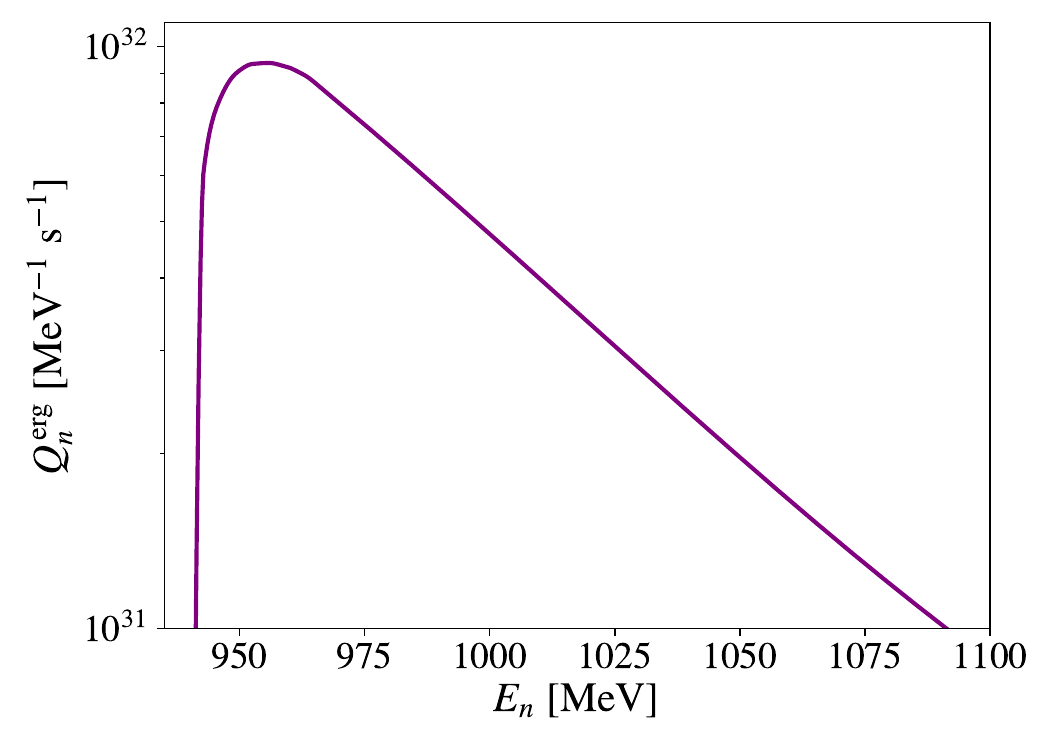}
    \caption{Differential neutron rate per unit energy and time reaching the ergosphere of Sgr~A* for the scenario with $a_\ast=0.94$. The resulting spectrum includes the effect of neutron decay along Kerr trajectories through the survival probability $P_{\rm surv}$.}
    \label{fig:Qsurv_kerr}
\end{figure}

We define the differential neutron rate  reaching the ergosphere, accounting for the survival probability of neutrons, as:
\begin{equation}
    Q_{n}^{\rm surv}=Q_{n}\times P_{\rm surv},
    \label{eq:Q_n_surv}
\end{equation}
where $Q_{n}$ is the total neutron production rate per unit energy, time, and volume introduced in Eq.~\eqref{eq:Q_n}. 
In Figure~\ref{fig:Qsurv_kerr} we present the result for the scenario with $a_\ast=0.94$, where $Q_{n}^{\rm surv}$ is integrated over the corresponding volume to obtain the total spectrum of neutrons that reach the ergosphere, $Q_{n}^{\rm erg}$.
We note that very similar results are obtained for the other scenarios considered in this work (i.e., $a_\ast=0.5$ and $a_\ast=0.7$), with only minor variations arising from the dependence of the neutron survival probability on the BH spin. 

As already mentioned, the region where the MPP can take place may extend beyond the ergosphere~\cite{Dhurandhar1984_I, Dhurandhar1984_II}, provided that the condition for the existence of negative-energy orbits is satisfied (see Section~\ref{subsec:MPP_theory} for details). 
Nonetheless, after explicitly computing the neutron population reaching radii larger than $R_{\rm S}$, we find that the gain is modest. In particular, extending the energy extraction region up to $5 \, R_{\rm S}$ increases the neutron rate only by a factor of $\sim 2$ at low energies, while the impact is even smaller at higher energies. This indicates that the dominant contribution arises from the innermost region, justifying our focus on the ergosphere, while adopting a conservative approach.  

\subsection{Accelerated protons through MPP}
\label{subsec:protons_MPP}

Owing to electromagnetic interactions, the MPP can achieve efficiencies exceeding $100\%$ even for relatively weak magnetic fields of order $\sim \mathrm{mG}$ \cite{1985JApA....6...85B, 1986ApJ...307...38P, Dadhich_2018}, and may reach values as high as $\sim 10^{12}$ under optimal conditions \cite{Tursunov_2020}. Here, we define the efficiency of the process as the ratio between the net extracted energy and the initial energy of the infalling particle, i.e.,

\begin{equation}
\eta=(E_3-E_1)/E_1,
\label{eq:eta}
\end{equation}
where the subscripts $1$ and $3$ refer to the initial and escaping particles, which in our scenario correspond to the neutron and proton, respectively.

The MPP can operate in different regimes — low, moderate, and ultra-high efficiency — depending on the charge configuration and magnetic field strength \cite{Tursunov_2019}. In the scenario considered here, where the charge of the escaping particle exceeds that of the infalling one ($q_3 > q_1$), the process operates in the ultra-high efficiency regime. In this case, from Eq.~\eqref{eq:eta}, the efficiency can be expressed as \cite{Tursunov_2019}:
\begin{equation}
    \eta_{\rm MPP}=\eta_{\rm PP}-\frac{q_3}{m_1}A_t,
\label{eq:etaMPP2}
\end{equation}
where $\eta_{\rm PP}=1/2 (\sqrt{R_{\rm S}/r}-1)$ denotes the efficiency of the original PP \cite{Penrose:1969pc, 1972ApJ...178..347B}, ranging from zero for non-rotating BHs to a maximum value of $\sim 20.7\%$ in the extremal Kerr limit,
and $A_t$ is given in Eq.~\eqref{waldpotentialt}. Note that the temporal component of the vector potential, $A_t$, becomes negative when the BH spin and magnetic field are aligned ($aB>0$). Since the second term in Eq.~\eqref{eq:etaMPP2} dominates due to the large charge-to-mass ratio, positive efficiencies are only obtained under this alignment condition for positively charged escaping particles ($q_3>0$). In our case, this condition is fulfilled, since we consider protons as the escaping particle and a positively rotating BH~\cite{dayem2026}, implying an aligned configuration between the spin and the magnetic field. This ensures that positive efficiencies can be achieved within our framework. 

Besides, assuming that the magnetic field in the small region where neutron decay occurs is well described by the Wald solution, we can substitute the expressions of Eqs.~\eqref{waldpotentialt} and \eqref{waldpotentialphi} for the electromagnetic potential into the previous equation to obtain the well-known expression for the MPP efficiency in the equatorial plane ($\theta=\pi/2$) recovering physical units:
\begin{equation}
    \eta_{\rm MPP}=\frac{1}{2}\left(\sqrt{\frac{R_{\rm S}}{r_{\rm split}}}-1\right)+\mathcal{B} \left( 1-\frac{R_{\rm S}}{2r_{\rm split}}\right),
    \label{eq:eta_MPP}
\end{equation}
with:
\begin{equation}
    \mathcal{B}=\frac{q_{3}BMGa_\ast}{m_{1}c^4}.
\end{equation}
Here, $B$ denotes the strength of the uniform magnetic field, $a_\ast$ is the dimensionless spin parameter, and $r_{\rm split}$ is the splitting point, i.e., the radius at which the particle decay occurs. The first term is purely geometrical and corresponds to the original PP, while the second term accounts for the contribution of the magnetic field.

A key aspect of this mechanism is the location of the splitting point. If the decay occurs too close to the event horizon (e.g., at $r = r_+$), both fragments may be captured by the BH due to the extreme gravitational redshift. On the other hand, if the decay takes place too far from the horizon, neither fragment may be captured, preventing the energy extraction mechanism from operating. Therefore, an optimal configuration is achieved when the splitting occurs sufficiently close to the outer event horizon — typically at $\sim R_{\rm S}$ \cite{Tursunov_2020} — such that one fragment falls into the BH while the other escapes to infinity. In this study, we take $r_{\rm split} = R_{\rm S}$, thus, the process can efficiently produce a flux of accelerated protons where all the efficiency contribution comes from the second term of Eq.~\eqref{eq:eta_MPP}, i.e., the magnetic part.

\begin{figure}[h!]
\centering
\includegraphics[width=1.0\linewidth]{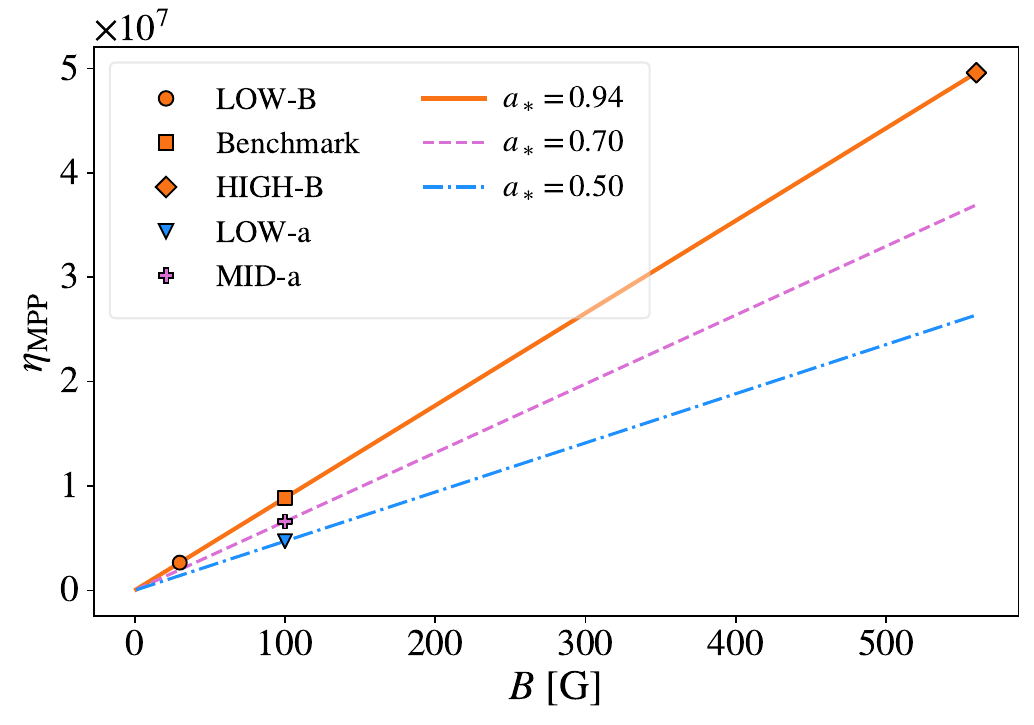}
\caption{MPP efficiency ($\eta_{\rm MPP}$) as a function of the magnetic field ($B$) for different values of the dimensionless spin parameter of Sgr~A*, and assuming a BH mass of $M = 4 \times 10^{6}\, M_{\odot}$. We consider protons as the accelerated escaping particles and take the splitting point at $r_{\rm split}= R_{\rm S}$. We also represent the different scenarios considered in this work, listed in Table \ref{tab:scenarios}.}
\label{etaB}
\end{figure}

Figure~\ref{etaB} shows the dependence of the MPP efficiency on the magnetic field strength and the BH spin. As expected, $\eta_{\rm MPP}$ increases with both $B$ and $a_\ast$, indicating that higher efficiencies are achieved in environments with stronger magnetic fields and more rapidly rotating BHs.\footnote{Nevertheless, for a maximally rotating (extremal) BH with $a_\ast=1$, the efficiency reduces to that of the original PP, reaching a maximum of $\sim 20.7\%$.} For the specific case of Sgr~A*, we present the results within the range of magnetic field strengths and spin values inferred from the most recent EHT observations (see discussion in Section~\ref{subsec:MAD}), namely $B \sim 30 - 560 \; \rm G$ depending on the distance to the event horizon, and a relatively high spin of $0.5$ or $0.94$, with $a_\ast=0.94$ being favored \cite{2024ETH_VIII}.
To quantify these trends, we evaluate the efficiency of the process for the different scenarios considered in this work (see Table~\ref{tab:scenarios}). The resulting efficiency values for these configurations are summarized in Table~\ref{tab:efficiency_vals}.

\begin{table}[h]
\centering
\begin{tabular}{lccc}
\hline
Model & $\log_{10}(\eta_{\rm MPP})$ & $E_p^{\max}$ [PeV] & $\log_{10}\left(P_{\mathrm{kin}}/\mathrm{erg\,s^{-1}}\right)$ \\
\hline
\texttt{LOW-B}     & $6.42$ & $2.93$  & $37.45$ \\
\texttt{BENCHMARK} & $6.95$ & $9.75$  & $37.97$ \\
\texttt{HIGH-B}    & $7.70$ & $54.56$ & $38.72$ \\
\texttt{LOW-a}     & $6.67$ & $5.18$  & $37.69$ \\
\texttt{MID-a}     & $6.82$ & $7.26$  & $37.84$ \\
\hline
\end{tabular}
\caption{MPP efficiency $\eta_{\rm MPP}$, maximum proton energy $E_p^{\max}$, and jet kinetic power $P_{\mathrm{kin}}$ for the different scenarios considered in this work summarized in Table~\ref{tab:scenarios}. The MPP efficiency and kinetic power are expressed in logarithmic form as $\log_{10}(\eta_{\rm MPP})$ and $\log_{10}\left(P_{\mathrm{kin}}/\mathrm{erg\,s^{-1}}\right)$, respectively.}
\label{tab:efficiency_vals}
\end{table}

Using the MPP efficiencies derived above, we can now compute the spectrum of accelerated protons. From Eq.~\eqref{eq:eta}, the energy of the escaping protons produced after neutron beta decay is given by:
\begin{equation}
    E_p= (\eta_{\rm MPP}+1)E_n,
\label{p_energy}
\end{equation}
where we assume, as a good approximation, that the protons inherit almost all the energy of the parent neutrons after their decay, as demonstrated in Appendix~\ref{app:beta_decay}.

\begin{widetext}
    
\begin{figure}[h]
\begin{center}
		\begin{subfigure}{0.49\textwidth}
			\includegraphics[width=\textwidth]{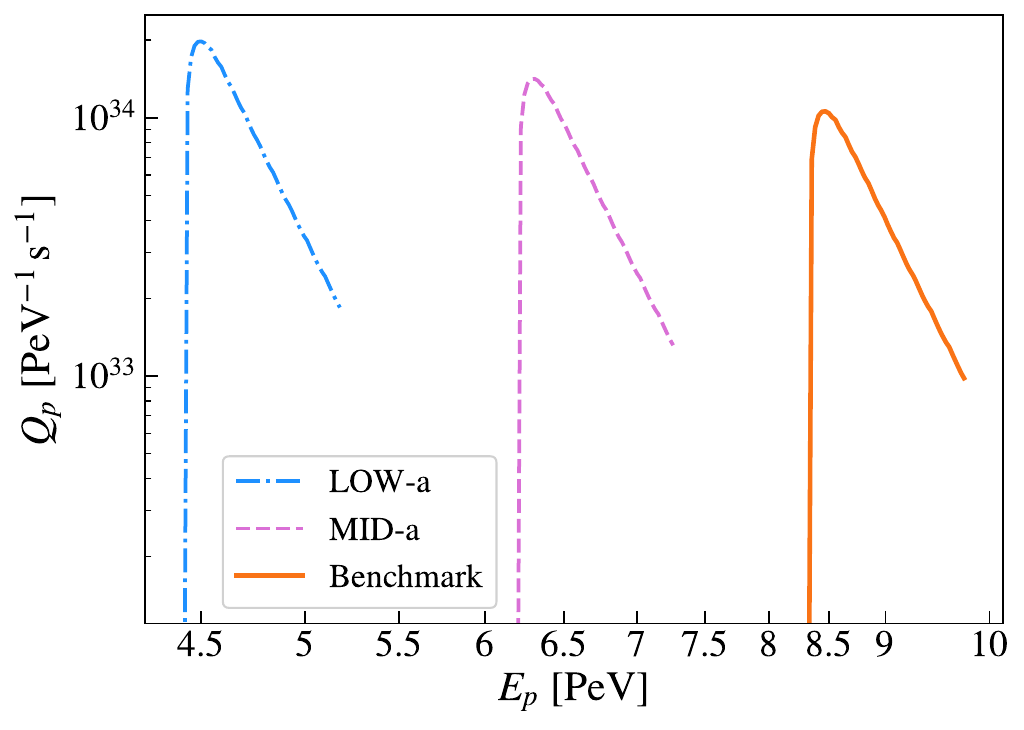}
		\end{subfigure}
		\hspace{0.01cm}
		  \begin{subfigure}{0.49\textwidth}
		      	\includegraphics[width=\textwidth]{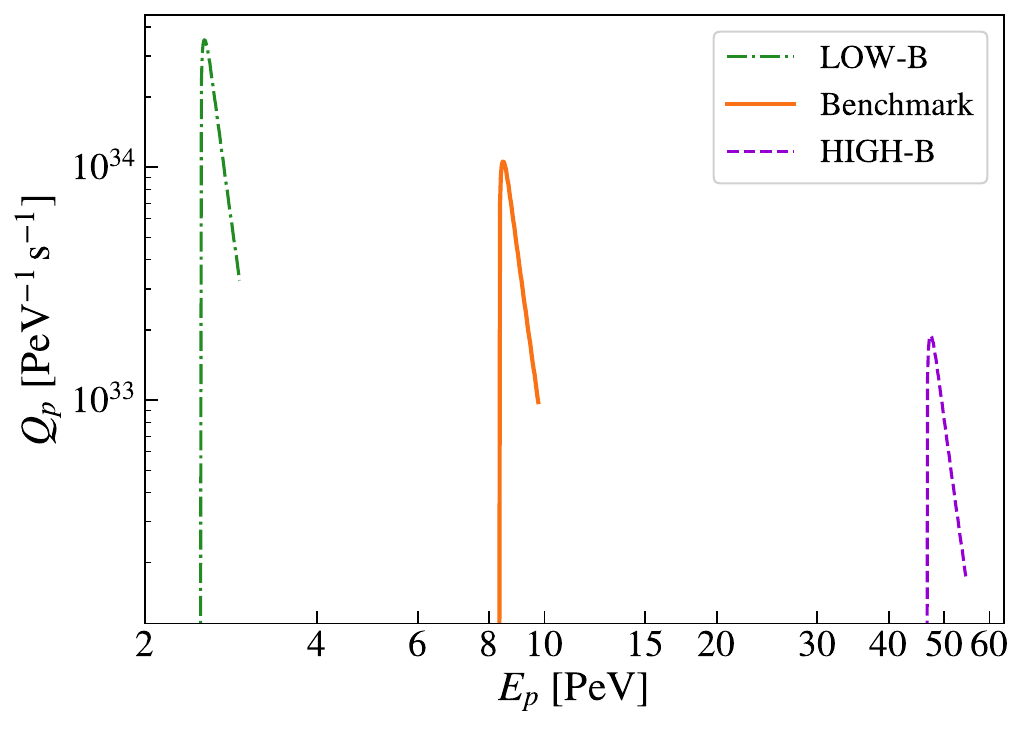}
		  \end{subfigure}
        \caption{Differential proton rate per unit energy and time of the resulting accelerated protons via MPP in Sgr~A*. The left panel displays the cases corresponding to the \texttt{Benchmark}, \texttt{LOW-a}, and \texttt{MID-a} scenarios, while the right panel shows the results for the \texttt{Benchmark}, \texttt{LOW-B}, and \texttt{HIGH-B} cases; see Tables \ref{tab:scenarios} and \ref{tab:efficiency_vals}. This comparison illustrates the impact of varying the BH spin and magnetic field strength on the resulting proton flux.
        }
        \label{amplified_p}
		
\end{center}
\end{figure}
\end{widetext}

Figure~\ref{amplified_p} shows the differential production rate of accelerated protons per unit energy and time, $ Q_{p}=dN_{p} / dE_{p} \,dt$,  as a function of the proton energy. As expected, the proton energy increases with both the magnetic field strength and the BH spin. In contrast, $Q_p$ exhibits the opposite trend, decreasing as these parameters increase.
The left panel displays the cases with fixed $B=100 \; \rm G$ and varying spin. While $E_p$ increases with $a_\ast$, the production rate $Q_p$ is higher for lower spin values and decreases for rapidly rotating BHs. This behavior can be understood in terms of particle dynamics in a rotating spacetime: the BH rotation tends to trap particles in co-rotation, delaying their infall into the event horizon. As a result, for higher spin values, particles remain confined for longer times before being captured, leading to a reduction in the effective proton production rate.
A similar trend is observed when varying the magnetic field strength. In the right panel of Figure~\ref{amplified_p}, we show cases with fixed spin $a_\ast = 0.94$ and different values of $B$. Again, $E_p$ increases with $B$, while $Q_p$ decreases. Stronger magnetic fields enhance particle confinement near the BH, increasing the residence time prior to capture and consequently reducing the proton production rate.

The kinetic power of the jet, composed of accelerated protons via MPP, can be estimated as:
\begin{equation}
    P_{\rm kin}=\int_{E_{\rm min}}^{E_{\rm max}} Q_p(E,t) \hspace{1mm} E_p \hspace{1mm} dE_p.
\end{equation}
Based on the results presented above, the $P_{\rm kin}$ values for the different models considered in this work are listed in Table~\ref{tab:efficiency_vals}.

Although relativistic jets are a common feature of many accreting BHs, there is currently no conclusive observational evidence for the presence of a persistent jet in Sgr~A*. Nevertheless, several radio, X-ray, and kinematic structures have been interpreted as possible signatures of a weak collimated outflow \cite{falcke2001, Li_2013, yusef2020}, while others have argued that these features can be equally well explained by winds or by the complex morphology of the accretion flow itself (e.g., \cite{Yuan2014, Wang_2013}). Consequently, the existence of a jet in Sgr~A* remains an open question. If such a jet is present, it is likely to be intrinsically weak, compact, and radiatively inefficient \cite{Markoff2007}, making its direct detection particularly challenging. In this context, the kinetic powers derived in this work should be regarded as theoretical estimates for a proton-dominated outflow powered by the MPP, rather than as evidence for an observationally confirmed jet.

\section{Predicted high-energy emission signatures}
\label{sec:obs_signatures}

HECRs and UHECRs, with energies exceeding $\sim 10^{15} \,\rm eV$ and $10^{18} \, \rm eV$, respectively, remain among the most challenging phenomena in high-energy astrophysics, as their origin and acceleration mechanisms are still not fully understood. The extension of the cosmic-ray (CR) spectrum beyond PeV energies points to the existence of so-called PeVatrons, astrophysical sources capable of accelerating particles up to at least $\sim 10^{15} \, \rm eV$~\cite{wilhelmi2024huntpevatronsoriginenergetic}. Although a definitive identification remains elusive, a variety of candidates have been proposed, including supernova remnants (e.g.,~\cite{Ackermann_2013,Jouvin_2017}), microquasars (e.g.,~\cite{kaci2025microquasarsmajorcontributorsgalactic}), or Sgr~A* (e.g.,~\cite{2016hess, Fujita_2017, Guo_2017, Rodr_guez_Ram_rez_2019}), among others. 
In this Section, we explore the potential high-energy observational signatures of the MPP in Sgr~A*. 

\subsection{MPP as a powerful high-energy CR injector}
\label{subsec:cosmic_rays}

Previous studies have suggested that the ultra-high efficiency regime of the MPP could provide a viable mechanism for accelerating CRs to extreme energies~\cite{Tursunov_2020, Turs2022}.
For instance, stellar-mass BHs could accelerate protons up to $10^{16} \, \rm eV$, while SMBHs might reach energies beyond $10^{20} \, \rm eV$~\cite{Turs2022}. In the specific case of Sgr~A*, previous estimates indicate that proton energies as high as $\sim 10^{15.5} \, \rm eV$ can be achieved through the MPP~\cite{Tursunov_2020}, remarkably close to the \textit{knee} of the CR spectrum. 
This spectral feature is commonly interpreted as marking the maximum energy to which Galactic sources accelerate protons, yet its physical origin remains unexplained despite decades of study~\cite{H_randel_2004}.

\begin{figure}[h!]
    \centering
    \includegraphics[width=1.0\linewidth]{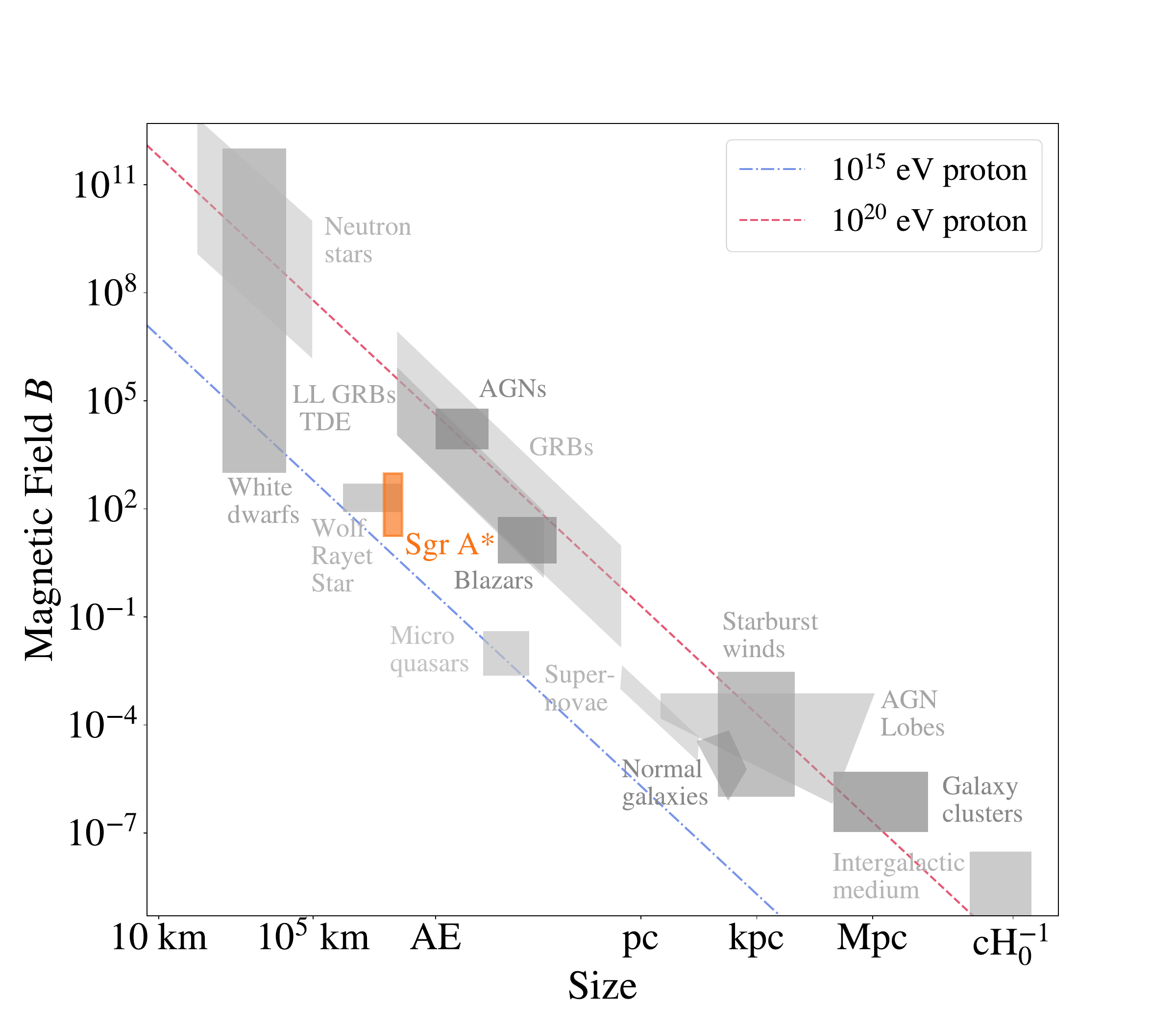}
    \caption{Magnetic field strength versus characteristic size for potential HECR and UHECR astrophysical accelerators, i.e., the so-called Hillas diagram~\cite{1984Hillas}. The Sgr~A* parameter space explored in this work is shown as a colored region together with a compilation of known HECR and UHECR candidate accelerators taken from Ref.~\cite{gernot_maier_2022_6037985}. The region spans the range of magnetic field strengths considered here, while its characteristic size remains unchanged for different BH spin values.  The diagonal lines indicate the Hillas confinement condition for protons with energies of $10^{15} \,\rm eV$ and $10^{20} \,\rm eV$. Our results place Sgr~A* in the region associated with Galactic PeVatron candidates.}
    \label{fig:Hplot}
\end{figure}

To assess the astrophysical significance of the MPP in Sgr~A*, first, we place our results alongside established HECR and UHECR source candidates taken from Ref.~\cite{gernot_maier_2022_6037985} in the so-called Hillas diagram, that compares their characteristic sizes, magnetic field strengths, and maximum proton energies~\cite{1984Hillas}, as illustrated in Figure~\ref{fig:Hplot}.  
The Hillas plot provides a useful qualitative framework to evaluate the accelerator capability of our source in the context of other candidate PeVatrons and UHECR emitters, considering the maximum proton energies obtained (Table~\ref{tab:efficiency_vals}).
The colored region shown in Figure~\ref{fig:Hplot} for Sgr~A* correspond to the \texttt{LOW-B}, \texttt{Benchmark}, and \texttt{HIGH-B} scenarios (see Table~\ref{tab:scenarios}), which span the range of magnetic field strengths explored in this work.
Note that we do not include the scenarios varying the spin parameter, since the acceleration region is identified with the equatorial ergosphere, whose size is independent of spin in the equatorial plane. 
Consequently, varying the magnetic field strength changes the vertical extent of the Sgr~A* region, whereas varying the spin does not shift its position along the size axis.  
Our results show that the particle emission expected from Sgr~A* through the MPP falls within the region occupied by Galactic PeVatrons in the Hillas diagram, highlighting its potential as an efficient high-energy particle accelerator.

\subsection{Gamma-ray and neutrino emission}
\label{subsec:gammas}

Recent observational and theoretical developments challenge the paradigm that CRs are accelerated up to the knee primarily by supernova remnants, pointing instead to the need for alternative sources capable of explaining current high-energy and UHECR observations~\cite{wilhelmi2024huntpevatronsoriginenergetic, Cardillo_2023}. Although several classes of the so-called PeVatron candidates have been proposed, no conclusive identification of their origin has yet been achieved.
In this context, UHE gamma rays and neutrinos constitute powerful messengers for the identification of PeVatrons, thanks to the unprecedented observational capabilities of recent UHE facilities~\cite{HAWC:2013kjc, CTAConsortium:2017dvg, LHAASO:2019qtb, Takita:2023pgn}.  
These secondary messengers are produced in interactions of the UHECRs with the surrounding gas.
In particular, proton–proton ($p-p$) interactions transfer a significant fraction of the primary proton energy to secondary gamma rays and neutrinos through the production of pions, where gamma rays originate from the decay of neutral pions and neutrinos from the decay of charged pions. Therefore, the detection of UHE gamma rays or neutrinos from an astrophysical source provides strong evidence that the parent protons were accelerated to even higher energies.

In the case of Sgr~A*, the surrounding environment offers a natural target for such interactions. The GC hosts the so-called Central Molecular Zone (CMZ), a dense disk-like region rich in molecular gas extending over the inner $\sim 200 \, \rm pc$, with a height of $\sim 75 \, \rm pc$~\cite{1996Morris, battersby20253dcmzicentral}. Protons accelerated via the MPP and escaping from the vicinity of the BH would reach this region, where they efficiently interact with the ambient gas, producing gamma rays and neutrinos. This scenario is further supported by the expected geometry of the system: observational evidence suggests that the spin axis of Sgr~A* is not aligned with our line of sight and is likely tilted with respect to the Galactic plane~\cite{Event_Horizon_Telescope_Collaboration_2022}, implying that any outflow or particle escape is not strictly perpendicular to the plane.  As a result, a significant fraction of the accelerated protons can intercept the CMZ, enhancing the probability of hadronic interactions. In contrast, if the system were oriented such that particles were ejected along a direction perpendicular to the Galactic plane, their interaction with the CMZ would be strongly suppressed. Such configuration may be relevant for other SMBH, but are not expected to apply in the case of Sgr~A* considered here.

Observations of the GC region performed by H.E.S.S.~\cite{2016hess}, the Major Atmospheric Gamma-ray Imaging Cherenkov Telescope (MAGIC;~\cite{2020Acciari}) and the Very Energetic Radiation Imaging Telescope Array System (VERITAS;~\cite{Adams_2021}) have revealed the existence of a point-like source, namely HESS J1745-290, whose position is consistent with that of Sgr~A*~\cite{2016hess}. However, the limited angular resolution does not allow an unambiguous identification of its origin, and several counterparts have been proposed, including the pulsar wind nebula G359.95$-$0.04~\cite{2006Wang}, Sgr~A* itself (e.g.~\cite{Atoyan_2004, Fujita_2017, Rodr_guez_Ram_rez_2019}), and a central spike of annihilating dark matter (e.g.~\cite{Cembranos_2012}).

In addition to this compact source, H.E.S.S. has detected a VHE ($> 100 \, \rm TeV$) diffuse gamma-ray emission extending over the central $200 \, \rm pc$ of our Galaxy~\cite{2018hess}. This emission closely follows the dense gas distribution of the CMZ, suggesting a hadronic origin where UHECRs injected at the very GC interact with the ambient gas through $p-p$ collisions \cite{2016hess}. Remarkably, the centroid of the diffuse emission coincides with the location of HESS J1745-290, while the inferred CR density profile points to the presence of a persistent Galactic PeVatron operating near the dynamical center of the Milky Way. If the emission from HESS J1745-290 is predominantly hadronic, it would naturally imply the existence of a powerful source of relativistic protons in the immediate vicinity of the GC~\cite{2016hess}. In this scenario, both the diffuse and point-like gamma-ray components could originate from interactions of HECRs injected by a common central accelerator. Consequently, if Sgr~A* is indeed responsible for powering HESS J1745-290, it may also contribute to the population of PeV CRs required to explain the diffuse gamma-ray emission observed throughout the CMZ.

Motivated by these observations, we compute the expected gamma-ray emission produced by the interactions of protons accelerated via MPP in Sgr~A* with the CMZ gas, 
assuming that the escaping protons are emitted along the BH spin axis. Since this axis is inferred to be tilted with respect to the Galactic plane~\cite{Event_Horizon_Telescope_Collaboration_2022}, the resulting geometry naturally favors an efficient overlap between the accelerated particles and the CMZ, thus maximizing the probability of hadronic interactions.
Taking the differential proton injection rate per unit energy and time, $Q_{p}$, obtained from the MPP acceleration process (see Section~\ref{subsec:protons_MPP}), we assume that protons propagate diffusively through the surrounding medium before interacting with the gas of the CMZ. Under the assumptions of steady-state transport, isotropic diffusion, negligible energy losses over the propagation scale, and a spatially uniform diffusion coefficient, the stationary solution of the CR transport equation for a point-like source yields the differential proton flux at a characteristic distance $r_{\rm CMZ}$:

\begin{equation}
    \frac{dn}{dE_p} =
    \frac{Q_p(E_p)}{4 \pi  D(E_p) r_{\rm CMZ}},
\end{equation}
where $D(E_p)$ is the diffusion coefficient that characterizes the movement of charged particles in the Galaxy, 
and we take $r_{\rm CMZ}= 200 \, \rm pc$. 
Note that we assume that the accelerated protons propagate purely diffusively and with the same diffusion coefficient as inferred from local CR measurements~\cite{De_La_Torre_Luque_2021, DeLaTorreLuque2024DRAGON2Antiprotons}. 
However, the transport properties depend sensitively on the environment, local magnetic field configuration and turbulence level~\cite{Lazarian2023CRPropagation, EvoliYan2014}, and may be significantly different to those inferred from CR data. 
In particular, we assume that CR propagation is governed by Kolmogorov-type turbulence, leading to a diffusion coefficient given by $D(E_p)\sim D_{0}\times (E_{p}/E_{0})^{\delta} \; \rm cm^{2} \,s^{-1}$, with $D_{0}=4.53 \times 10^{28} \; \rm cm^{2} \,s^{-1}$, $E_{0}= 4 \, \rm GeV$ and $\delta=1/3$~\cite{Evoli_2018,Luque_2025}.

The gamma-ray production rate per unit of energy, time and volume is then derived from the interaction of the propagated proton distribution with the differential cross-section of $p-p$ interactions, for which we adopt the state-of-the-art AAfrag cross-sections~\cite{Kachelriess2019AAfrag}:
\begin{equation}
    Q_\gamma(E) =n_H\int dE_p\,\frac{dn}{dE_p(E_p)}\frac{d\sigma(E_p,E)}{dE},
\end{equation}
with $n_{\rm H}$ being the number density of the hydrogen gas, taken here to be uniform and $n_{\rm H}=200  \; \rm cm^{-3}$ in average for the whole CMZ~\cite{Gabici_2022}.

Finally, the observable gamma-ray flux at Earth is:
\begin{equation}
\begin{split}
    \frac{d\phi}{dE} = 
\frac{L_\gamma(E)}{4\pi d^2} = 
\frac{n_H V}{4\pi d^2}
\int dE_p
\,
\frac{Q_p(E_p)}{D(E_p) r_{\rm CMZ}} 
\frac{d\sigma(E_p,E)}{dE},
\end{split}
\end{equation}
where we take $d=8.3 \,\mathrm{kpc}$ as the distance to the GC~\cite{2021Gravity} and $L_{\gamma}(E)$ is the total gamma-ray luminosity obtained by accounting for the interaction volume $V$ of the CMZ as $L_{\gamma}(E)=V \,Q_\gamma(E)$.

\begin{widetext}
    
\begin{figure}[h]
\begin{center}
		\begin{subfigure}{0.49\textwidth}
			\includegraphics[width=\textwidth]{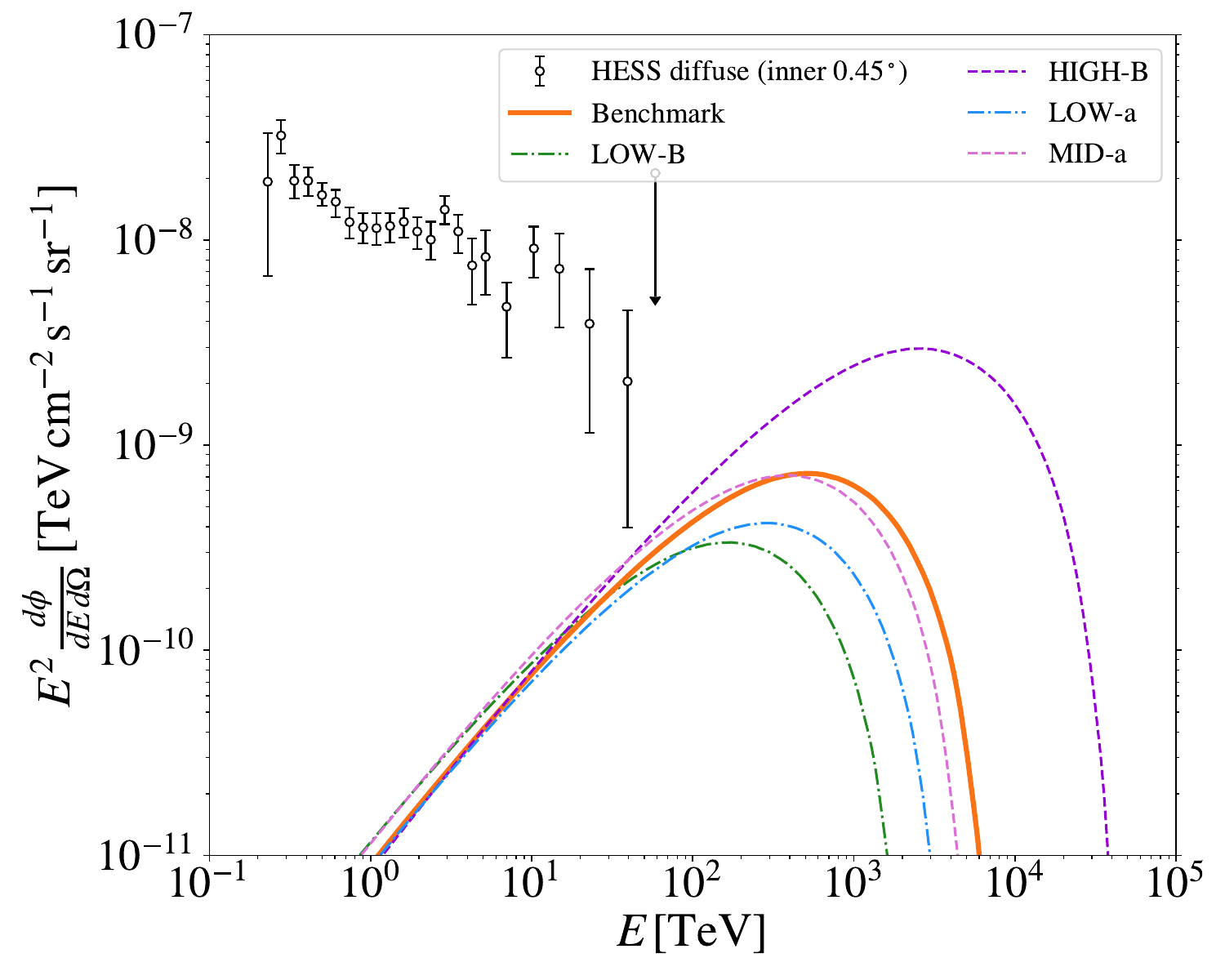}
			
		\end{subfigure}
		\hspace{0.01cm}
		  \begin{subfigure}{0.49\textwidth}
		      	\includegraphics[width=\textwidth]{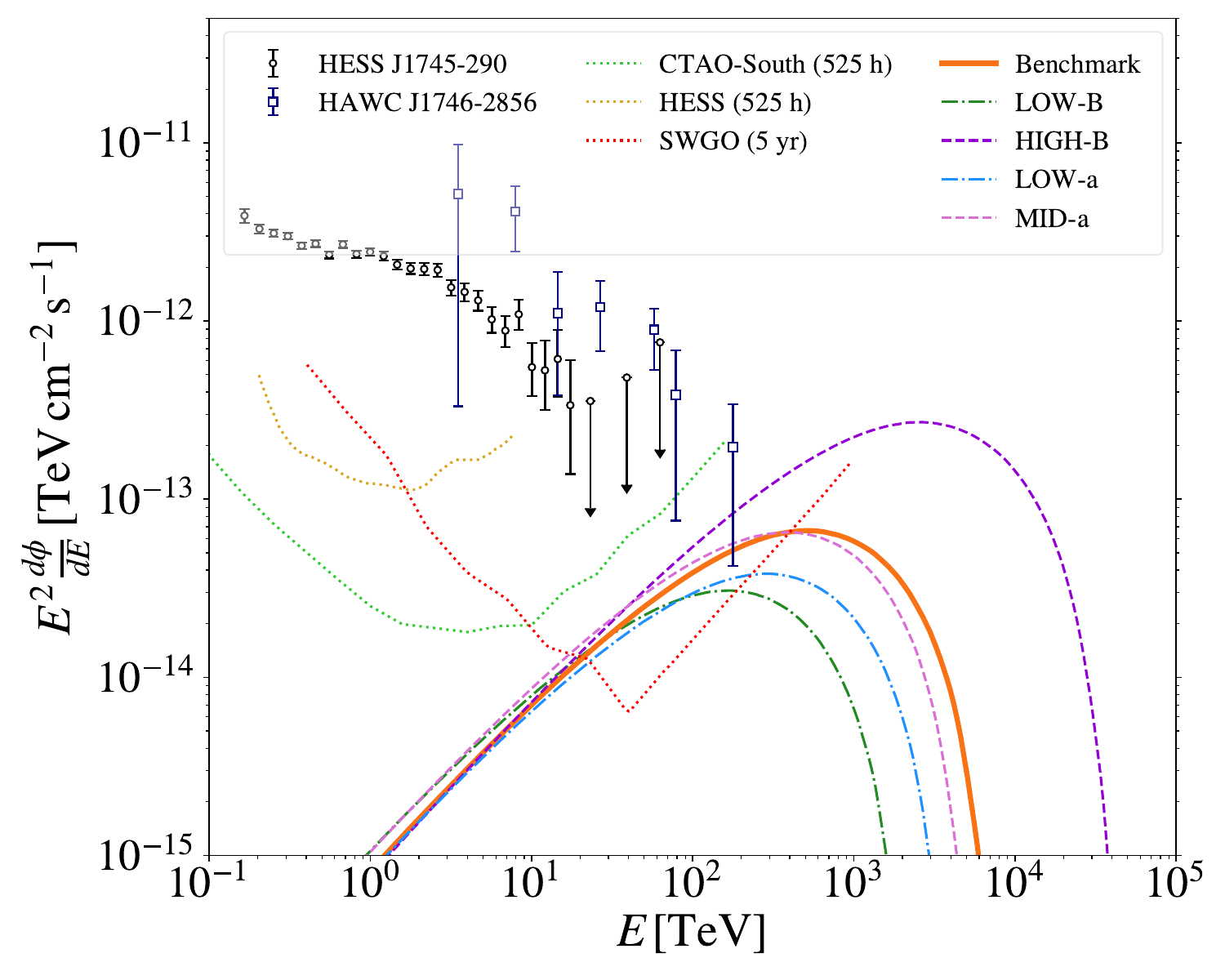}
			
		  \end{subfigure}
        \caption{Predicted gamma-ray fluxes from MPP-accelerated protons in Sgr~A* for the different scenarios considered in this work. The left panel shows the gamma-ray flux per unit solid angle, alongside the VHE diffuse emission observed by H.E.S.S. in the inner $0.45^\circ$ around the GC region~\cite{2018hess}. The right panel displays the corresponding fluxes computed within a central region of radius $10 \; \rm pc$ ($\sim 0.1^\circ$), together with the central point-like sources HESS J1745-290~\cite{2016hess} and HAWC J1746-2856~\cite{albert2024observationgalacticcenterpevatron}. Sensitivity curves of current and future gamma-ray instruments, namely H.E.S.S.~\cite{holler2015observationscrabnebulahess}, CTAO-South\footnote{\href{https://www.ctao.org/es/for-scientists/performance/}{https://www.ctao.org/es/for-scientists/performance/}} and SWGO~\cite{swgocollaboration2025scienceprospectssouthernwidefield}, are included to assess the observational prospects of the predicted emission (see Section~\ref{subsec:obs_prospects} for details).}
        \label{fig:gammas_CMZ_sgr}
		
\end{center}
\end{figure}
\end{widetext}

We show the gamma-ray flux obtained for the different scenarios considered in this work in Figure~\ref{fig:gammas_CMZ_sgr} (left panel). We compare our predictions with the diffuse gamma-ray emission measured by H.E.S.S. in the inner $0.45^\circ$ region around the GC, as the diffuse gamma-ray flux per unit solid angle was found to be approximately uniform across the CMZ, resulting in a spectrum consistent with that measured over the larger Galactic ridge region~\cite{2018hess}.
To enable a direct comparison with the H.E.S.S. observation, we express our predictions as a flux per unit solid angle by dividing the CMZ-integrated flux by the solid angle subtended by a spherical CMZ of radius $r_{\rm CMZ}=200 \,\mathrm{pc}$ at a distance of $8.3 \, \mathrm{kpc}$. The resulting quantity represents the average surface brightness of the CMZ and is shown in the same units as the H.E.S.S. measurements.
Overall, our predicted gamma-ray fluxes remain below, yet very close, to the observed diffuse VHE emission, suggesting that the MPP operating in Sgr~A* could provide a non-negligible contribution to the observed gamma-ray signal. 
Although additional processes are likely to contribute to the observed diffuse gamma-ray emission, our predicted fluxes indicate that this mechanism may constitute a significant source of high-energy particles in the GC region.
In this context, our results support the possibility that Sgr~A*, through the action of the MPP, contributes to the population of high-energy protons responsible for the diffuse gamma-ray emission. 
Furthermore, the maximum proton energies attained in our scenarios reach the PeV range inferred from observations of the GC, indicating that the MPP may represent a viable acceleration mechanism for powering a Galactic PeVatron associated with Sgr~A*.

In addition to the diffuse emission from the CMZ, H.E.S.S. observations revealed a point-like VHE gamma-ray source within the inner $\sim 10 \; \rm pc$ of the GC region, HESS J1745-290, exhibiting a hard power-law spectrum extending up to tens of TeV without evidence for a cutoff~\cite{2016hess}, whose spectral behavior is considered one of the characteristic signatures of a PeVatron capable of accelerating particles up to PeV energies. Based on the observed CR density profile and the localization of the emission, the H.E.S.S. collaboration argued that Sgr~A* could be linked to the accelerator powering this central PeVatron. Moreover, the HAWC Observatory has reported a point-like source, HAWC J1746-2856,  spatially coincident with HESS J1745-290 and with no evidence of a spectral cutoff up to 100 TeV, also confirming the existence of a GC PeVatron~\cite{albert2024observationgalacticcenterpevatron}.
Therefore, we also compute the gamma-ray flux expected from MPP-accelerated protons interacting within the innermost $10 \; \rm pc$ ($\sim 0.1^\circ$) region surrounding the GC. For the gas distribution, we follow the model presented in Ref.~\cite{Linden_2012}, in which the gas number density averaged over solid angle varies with radius, taking values of $n_{\rm H}\sim 2\times10^{3} \; \rm cm^{-3}$ between $0.1 - 1 \; \rm pc$ and $3 - 10 \; \rm pc$, and $n_{\rm H}\sim 8\times10^{4} \; \rm cm^{-3}$ between $1 - 3 \; \rm pc$. 
Our results are displayed in Figure~\ref{fig:gammas_CMZ_sgr} (right panel), along with the measured spectra of HESS J1745-290 and HAWC J1746-2856. All the predicted fluxes remain below the observed emission and are consistent with the current upper limits, implying that none of the considered scenarios can be excluded by existing observations. Yet, our results suggest that MPP-accelerated protons from Sgr~A* could provide a relevant contribution to the gamma-ray emission of HESS J1745-290 and HAWC J1746-2856. 

\begin{figure}[h!]
    \centering
    \includegraphics[width=1.0\linewidth]{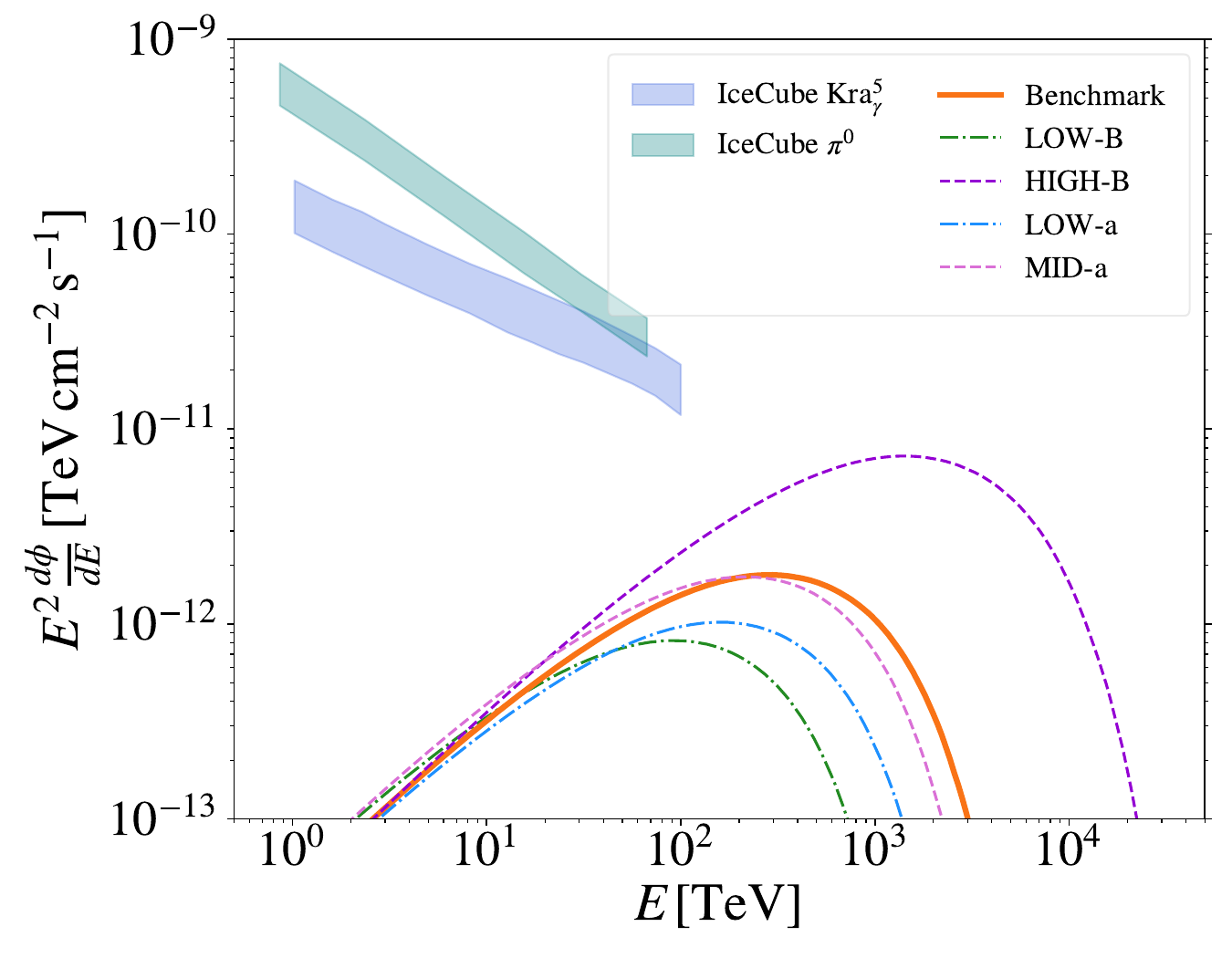}
    \caption{Predicted neutrino flux from MPP-accelerated protons in Sgr~A* for each scenario considered in this work (Table~\ref{tab:scenarios}), compared with IceCube best-fit models for the GC region~\cite{2023icecube}, including the $\rm KRA_\gamma^5$ and $\pi^{0}$-decay models~\cite{Gaggero_2015}.}
    \label{fig:neutrinos}
\end{figure}

In addition to gamma rays, $p-p$ interactions also produce high-energy neutrinos through the decay of charged pions and their secondary muons. Interestingly, the detection of VHE neutrinos would be the smoking gun for hadronic CR acceleration~\cite{Halzen_2002, curtisginsberg2026ultrahighenergygammaraysourcesneed}. We compute the expected neutrino flux following the same formalism adopted above for gamma rays. Starting from the differential proton injection rate per unit energy and time, $Q_{p}$, the neutrino production rate is obtained by replacing the differential gamma-ray production cross-section with the corresponding neutrino yield from $p-p$ interactions. The resulting neutrino luminosity is then integrated over the CMZ volume and converted into an observable flux at Earth adopting the distance to the GC. We display the neutrino flux results for the different scenarios considered in this work in Figure~\ref{fig:neutrinos}. 

We compare our predictions with the high-energy Galactic diffuse neutrino emission inferred by the IceCube Neutrino Observatory~\cite{2023icecube}, using the $\rm KRA_\gamma^5$ and $\pi^{0}$-decay templates~\cite{Gaggero_2015}. These measurements have recently been updated in Ref.~\cite{abbasi2026highenergyneutrinoemissionmilky}, and are in good agreement with the previous results. While IceCube has reported evidence for a diffuse high-energy neutrino signal from the Galactic plane, its angular resolution and sensitivity are not yet sufficient to isolate the contribution from the GC. The observed emission is broadly consistent interactions of Galactic CRs with the interstellar medium, although its precise origin and the relative contribution of individual sources remain uncertain~\cite{DeLaTorreLuque:2025zsv,ANTARES:2025wvi, DeLaTorreLuque2022, DeLaTorreLuque2023PeVFrontier}.
Although our model alone is not expected to account for the total neutrino flux from the Galactic plane, these results suggest that the MPP may represent a viable source of high-energy neutrinos in the GC region.

\subsection{Future observational prospects and detectability}
\label{subsec:obs_prospects}
Although the results presented in the previous Section already suggest that MPP acceleration in Sgr~A* may have observable high-energy signatures in current instruments, in this Section we explore whether the predicted fluxes can be effectively probed with future observations by current and next-generation VHE observatories. 

To this end, in Figure~\ref{fig:gammas_CMZ_sgr} (right panel) we compare our predicted gamma-ray fluxes with the sensitivity curves of several instruments. We note, however, that these sensitivity curves are generally computed assuming both a generic power-law spectrum and a specific observational setup. 
Since the MPP predicts a characteristic non-power-law spectral shape, and the actual observing conditions may differ from those adopted in the published sensitivity estimates, the comparison presented here should be regarded as indicative of the expected detectability rather than as a precise sensitivity assessment. 
We show the 525-hour sensitivity of H.E.S.S.~\cite{holler2015observationscrabnebulahess}, finding that all the scenarios considered in this work remain below its detection threshold. While this is consistent with the fact that our predicted emission does not exceed the observed flux from the central point-like source HESS J1745-290, it also highlights the need for more sensitive observatories to further test the MPP mechanism. Hence, we also include the expected 525-hour sensitivity of the southern array of the upcoming Cherenkov Telescope Array Observatory (CTAO;~\cite{Gueta_2021}). This exposure time corresponds to the planned observing time of the GC Survey Key Science Project~\cite{CTAConsortium:2017dvg}. We do not show the sensitivity of CTAO-North since the GC will be primarily observed by the southern array. 
Although our predicted fluxes remain below the nominal CTAO-South sensitivity curve, the difference is only a factor of $\sim 2-3$ over the $10 - 100 \; \rm TeV$ range. Since sensitivity curves represent benchmark performance estimates derived for specific spectral and observational assumptions, such a modest separation should not be interpreted as ruling out detectability. A dedicated CTAO analysis, including simulations tailored to the specific spectral shape and spatial morphology predicted here, would be required for a robust assessment. 
Finally, we consider the Southern Wide-field Gamma-ray Observatory (SWGO;~\cite{albert2019sciencecasewidefieldofview}), a next-generation wide-field air-shower observatory designed to monitor the southern gamma-ray sky at energies extending to the TeV and PeV domains. We show its projected five-year sensitivity, while the curve for energies above $\sim 10^2 \, \rm TeV$ corresponds to an extrapolation performed in Ref.~\cite{swgocollaboration2025scienceprospectssouthernwidefield}. In this case, the prospects are considerably more favorable. Our results indicate that SWGO should be capable of probing all the scenarios that we are considering in this work. 
These findings highlight the crucial role that future gamma-ray observatories will play in testing the MPP mechanism operating in Sgr~A*, assessing its potential contribution to the VHE emission from the GC, and providing constraints on the spin and magnetic field strength of the central SMBH. 

Neutrino observations offer a complementary and potentially decisive probe of the hadronic nature of this scenario. In contrast to gamma rays, however, current neutrino measurements do not yet provide an isolated determination of the emission from the GC. Existing measurements mainly constrain the diffuse Galactic component through template analyses or all-sky Galactic fluxes, while dedicated searches for point-like or moderately extended emission from the inner Galaxy have so far yielded only upper limits~\cite{Albert_2020}. Nevertheless, the next generation of neutrino telescopes, particularly KM3NeT/ARCA~\cite{Aiello_2019km3net,2024km3net} and IceCube-Gen2~\cite{Aartsen_2021}, is expected to substantially improve the sensitivity to point-like sources and to neutrino emission from the inner Galaxy, enabling tests of whether a localized hadronic component associated with the CMZ, such as that predicted here, can be distinguished from the more extended diffuse Galactic emission. Complementary measurements by the Baikal Gigaton Volume Detector (Baikal-GVD; ~\cite{2022baikal}) could further contribute to establishing the origin of high-energy neutrinos from the inner Galaxy~\cite{gvdcollaboration2026constraintspointlikeastrophysicalsources}. Additionally, the proposed TRopIcal DEep-sea Neutrino Telescope (TRIDENT;~\cite{TRIDENT:2022hql}) may eventually provide additional sensitivity to the GC,  although robust source-specific sensitivities are not yet sufficiently established for a quantitative comparison.

\section{Discussion and conclusions}
\label{sec:conclusions}

In this work, we have explored the extraction of rotational energy from the SMBH at the center of our Galaxy, Sgr~A*, through the MPP~\cite{1985ApJ...290...12W}. This mechanism constitutes an electromagnetic extension of the original PP~\cite{Penrose:1969pc}, arising when the BH is embedded in an external magnetic field. In such environment, electromagnetic interactions modify the dynamics of charged particles within the ergosphere of the BH, substantially enhancing the efficiency of energy extraction and enabling the production of highly energetic escaping particles.

In particular, we study the ultra-high efficiency regime of the MPP, in which neutron beta decay takes place inside the ergosphere of Sgr~A*. We model the accretion flow with an ADAF model~\cite{Narayan_1994, Narayan_1995a, Narayan_1995b}, in which the high ion temperatures reached in the plasma allow an efficient neutron production through nuclear reactions.
As a novel aspect of this work, we estimate the neutron production spectrum in the accretion flow, directly from the underlying nuclear production process, rather than relying on thermally averaged reaction rates together with an assumed MB distribution, as adopted in previous studies (e.g., ~\cite{Guessoum_1999, Jean_2001, Oh_2024}). 
We then account for gravitational escape from the accretion flow and follow the trajectories of the surviving neutrons in the Kerr spacetime to determine the fraction that reaches the ergosphere before decaying, thereby identifying the population capable of undergoing the MPP.
Within our framework, where the external magnetic field is aligned with the BH spin axis, the energetically favoured outcome corresponds to electrons being captured by the BH, while protons escape to infinity with enhanced energies.
We derive the spectrum of accelerated protons by computing the MPP efficiency, assuming that the magnetic field can be treated as locally uniform over the small region where neutron decay occurs, so that the Wald solution~\cite{PhysRevD.10.1680} provides an appropriate description of the electromagnetic field.

Once accelerated, the escaping protons propagate into the environment surrounding Sgr~A*, where they undergo hadronic interactions with the dense gas of the CMZ producing gamma rays and neutrinos through $p-p$ collisions. This scenario is naturally supported by the expected geometry of the system, as observational evidence indicates that the spin axis of Sgr~A* is tilted with respect to the Galactic plane~\cite{Event_Horizon_Telescope_Collaboration_2022}, allowing a significant fraction of the accelerated protons to intercept the CMZ rather than escaping along a direction perpendicular to the disk. Based on this framework, we predict the gamma-ray and neutrino emission arising from MPP-accelerated protons. We also compare our results with current observations and the sensitivities of present and future high-energy observatories.

Our main findings can be summarized as follows:
\begin{itemize}
    \item The computed neutron spectrum broadly preserves the overall shape obtained under the conventional MB approximation, yet it exhibits significant differences in both its normalization and high-energy behavior. In particular, our calculation predicts a reduced population of high-energy neutrons compared to the MB approximation, with differences approaching one order of magnitude at the highest energies and growing with increasing production radius (see Figure~\ref{fig:inicial_neutron_rate}). This, in turn, leads to smaller gravitational escape losses (see Figure~\ref{fig:f_esc}). These results highlight the importance of deriving the neutron spectrum directly from the underlying nuclear reaction kinematics rather than assuming thermally averaged reaction rates. Since neutron production rates in ADAFs are also employed in a variety of astrophysical contexts beyond the MPP scenario considered here, this result may have broader implications for previous and future studies (see, e.g., Refs.~\cite{Guessoum_1999, Jean_2001}).
    
    \item Across all the scenarios explored here, the MPP is found to operate with remarkably high efficiencies, with values typically of $\eta_{\rm MPP}\sim10^{6}$--$10^{7}$ (see Table~\ref{tab:efficiency_vals} and Figure~\ref{etaB}). These efficiencies allow protons to be accelerated up to maximum energies of approximately $E_p^{\max}\sim1$--$10 \, \mathrm{PeV}$, depending on the assumed BH spin and magnetic field strength. Despite the variation among the different scenarios, these results consistently indicate that the MPP may operate as an extremely efficient mechanism for producing ultra-energetic protons in the vicinity of Sgr~A*.
    
    \item Our results strengthen the interpretation of Sgr~A* as a candidate Galactic PeVatron through a more refined treatment that consistently follows the neutron population from its production in the accretion flow to the subsequent MPP acceleration, while remaining consistent with previous studies suggesting that Sgr~A* would be able to accelerate protons up to PeV energies through the MPP~\cite{Tursunov_2020, Oh_2024}. 
    This interpretation is particularly relevant in light of the VHE observations of the GC. H.E.S.S. has detected both the point-like source HESS J1745-290~\cite{2016hess} and a diffuse TeV gamma-ray emission extending over the inner $\sim 200\,\rm pc$ of the Galaxy~\cite{2018hess}, while HAWC has also reported the spatially coincident source HAWC~J1746-2856~\cite{albert2024observationgalacticcenterpevatron}. In all the scenarios considered here, our predicted gamma-ray fluxes remain below the observed emission and the reported upper limits, indicating that none of the models is excluded by current observations (see Figure~\ref{fig:gammas_CMZ_sgr}). At the same time, the predicted fluxes lie sufficiently close to the measured values to suggest that MPP-accelerated protons from Sgr~A* could constitute a non-negligible contribution to the observed VHE emission from the GC.
    
    \item The predicted gamma-ray fluxes exhibit a characteristic spectral shape that differs from the simple power-law spectra commonly used to describe many VHE gamma-ray sources (e.g.,~\cite{hueyotlzahuantitla2017tevspectralenergydistribution, Abeysekara_2017, 2018hessGplane}). If confirmed observationally, this distinctive spectral signature could provide an observational fingerprint of an MPP contribution to the VHE emission of the GC region. 

    \item In the neutrino channel, the absence of a resolved high-energy neutrino signal from the GC prevents a comparison with observations at the same level as for gamma rays. Our predicted neutrino fluxes lie below the diffuse Galactic component inferred by IceCube~\cite{2023icecube} in all the scenarios explored, although the spectral shapes suggest that they may become comparable at higher energies (see Figure~\ref{fig:neutrinos}). These results indicate that MPP-accelerated protons in Sgr~A* could contribute to the high-energy neutrino emission from the GC region.
    
    \item Our study also highlights the promising prospects for testing the MPP with the next generation of high-energy observatories (see Figure~\ref{fig:gammas_CMZ_sgr}). For CTAO~\cite{Gueta_2021}, our predicted fluxes remain below the nominal sensitivity curve, but only by a factor of $2-3$ over part of the relevant energy range. However, given the dependence of this curve on the assumed source spectrum and observational setup, this modest separation does not preclude detectability. Besides, the predicted gamma-ray fluxes lie within the expected sensitivity of SWGO~\cite{swgocollaboration2025scienceprospectssouthernwidefield} for all the scenarios considered, opening the possibility of probing the MPP in the near future. Such a detection would provide the first observational evidence for the MPP mechanism that has so far remained purely theoretical, while simultaneously strengthening the case for Sgr~A* as a Galactic PeVatron. Moreover, the different MPP scenarios predict distinct gamma-ray fluxes, offering a complementary probe of the BH magnetic field strength and spin. On the neutrino side, future facilities such as KM3NeT/ARCA~\cite{Aiello_2019km3net,2024km3net} and IceCube-Gen2~\cite{Aartsen_2021} will significantly improve the sensitivity to Galactic sources and provide an independent test of this hadronic acceleration scenario. In particular, the detection of a corresponding VHE neutrino counterpart would constitute compelling evidence that the observed high-energy emission is powered by hadronic CR interactions.
\end{itemize}

These findings are subject to certain caveats that should be taken into account and addressed in future work. 
First, in computing the neutron production rate we consider only the dominant reaction, $p+\alpha \rightarrow p+n+{}^{3}\mathrm{He}$, while other reaction channels may also contribute. A fully self-consistent treatment would additionally require following the nuclear evolution of the accretion flow, including reactions that deplete the $\alpha$-particle population that acts as the neutron source. Moreover, the spectrum of accelerated protons is derived by evaluating the MPP efficiency at a representative decay location within the ergosphere, under the idealized assumption that neutron beta decay occurs sufficiently close to the outer event horizon.
While the ergosphere naturally defines the region where the MPP can operate, the efficiency depends sensitively on the exact decay location and the local spacetime conditions. Therefore, although our choice provides a well-motivated reference scenario in which the escaping and captured particles follow the energetically favoured outcome, decays occurring at different locations within the ergosphere could lead to lower energy extraction efficiencies, reducing both the proton yield and the maximum energies attained. However, this uncertainty is partly mitigated by our conservative choice of restricting the acceleration region to the ergosphere itself, neglecting any potential contribution from a broader region.

Another important caveat concerns the considerable uncertainty surrounding both the spin of Sgr~A* and the strength of its magnetic field. In this work, we have adopted the most recent values inferred from the EHT observations and modeling~\cite{2024ETH_VIII, Event_Horizon_Telescope_Collaboration_2022}. 
However, these parameters remain poorly constrained and depend to some extent on the underlying assumptions of the models used to interpret the observational data. Since both quantities directly influence the physical conditions and particle dynamics in the vicinity of the BH, their uncertainties propagate into our results. Nevertheless, we expect these uncertainties to be mitigated by the broad range of scenarios explored in this work (see Table~\ref{tab:scenarios}), which includes different combinations of the BH spin and magnetic field strength. Besides, we emphasize that the magnetic field around Sgr~A* is likely to possess a complex geometry. However, because the neutron decays considered here take place within a very small region of the ergosphere, the field can be treated as locally uniform and approximated by the Wald solution. Importantly, the applicability of the Wald solution does not require the magnetic field to be associated exclusively with a MAD configuration. Rather, it can be employed for a broader class of magnetic field geometries, provided that the field can be approximated as locally uniform in the region where particle decay occurs. Therefore, our results are not restricted to a strictly MAD state and remain applicable to more general magnetic field configurations around the BH. Under this assumption, the main uncertainty affecting our results arises from the magnetic field strength rather than from its local configuration.

Finally, the predicted gamma-ray and neutrino fluxes rely on a simplified description of CR transport in the GC. We adopt a standard steady-state diffusion scenario with isotropic propagation, negligible energy losses, and a spatially uniform diffusion coefficient inferred from local CR observations. However, the diffusion properties in the GC remain poorly constrained and may differ substantially because of its stronger magnetic fields, enhanced turbulence, and complex gas distribution~\cite{Lazarian2023CRPropagation, EvoliYan2014}. Our treatment implicitly assumes that the entire population of accelerated protons efficiently reaches the CMZ, whereas the actual fraction depends on the details of particle transport in the inner Galaxy. A more realistic description would therefore require dedicated CR propagation simulations tailored to the GC environment, which is beyond the scope of the present work and will be addressed elsewhere.

Future observational facilities will provide unprecedented opportunities to test models of particle acceleration and energy extraction in the immediate vicinity of Sgr~A*. In particular, the next generation of the Event Horizon Telescope (ngEHT;~\cite{Galison_2023}) is expected to deliver substantial improvements in angular resolution, sensitivity, and imaging fidelity over current horizon-scale observations, enabling more detailed studies of the dynamics and structure of the innermost accretion flow~\cite{Johnson_2023, Doeleman_2023}. In parallel, the Atacama Large Millimeter/submillimeter Array (ALMA;~\cite{Partnership_2015}), through its ongoing and future upgrades~\cite{Carpenter2019, Carpenter2022}, will continue to play a central role in global Very Long Baseline Interferometry (VLBI) networks, which combine widely separated radio telescopes to achieve an effective Earth-sized aperture and microarcsecond angular resolution. The enhanced capabilities of these facilities, together with coordinated multiwavelength observations, will provide tighter constraints on the magnetic field configuration, plasma properties, and particle acceleration processes operating near the event horizon. 

Such advances, combined with the next generation of gamma-ray and neutrino observatories, will provide an unprecedented opportunity to test the theoretical predictions of the MPP and, more generally, mechanisms of rotational energy extraction operating in the strong-gravity regime. By combining horizon-scale observations with multi-messenger measurements of high-energy particles, it will become possible to directly connect the plasma conditions in the immediate vicinity of the event horizon with their observable high-energy signatures. The framework developed here provides precisely such a connection, linking particle production and rotational-energy extraction near the BH to potentially observable signals. While the present work has focused on Sgr~A*, our methodology can be naturally extended to other magnetized BHs, including active galactic nuclei and other accreting SMBHs, opening a new avenue for exploring the role of BH energy extraction in high-energy astrophysics.

\begin{acknowledgments}
The authors would like to thank Sergio Hernández-Cadena for enlightening discussions.

The work of PDL, CFS, MASC and JZP was supported by the grant PID2024-155874NB-C21.
CFS's research has been partly carried out at the Institute for Computational Cosmology (ICC), whose hospitality is greatly appreciated, thanks to the CSIC iMOVE grant IMOVE24192.
MC acknowledge support from the Spanish Agencia Estatal de Investigaci\'on through the grants PID2024-155874NB-C22 and CEX2020-001007-S, funded by MCIN/AEI/10.13039/501100011033. MC also benefited from participating in COST Action COSMIC WISPers (CA21106), supported by COST (European Cooperation in Science and Technology). MC, PDL, CFS, VG, MASC and JZP acknowledge the MultiDark Network, ref. RED2022-134411-T. 
VG and JZP's work is supported by the project PID2022-139841NB-I00 funded by MICIU/AEI/10.13039/501100011033 and by ERDF/EU. VG also thanks the University San Pablo CEU research grant and the \textit{AstroLearning} group.
JZP's contribution to this work has been supported by \textit{FPI Severo Ochoa} PRE2021-099137 grant. The research of EMA has been supported by the grant CNS2022-135880. EMA also gratefully acknowledges the hospitality of the Niels Bohr Institute (NBI), where part of this research was carried out during a visit supported by the CSIC iMOVE programme (IMOVE24193). The research of MJR has been funded, in part, by the National Science Foundation under project number PHY-2309270. MJR also wants to thank the Mitchell Family Foundation for hosting her during the Cook’s Branch workshop, where some of the research was carried out.
PDL is supported by Ramón y Cajal RYC2024-048445-I grant, which is funded by MCIU/AEI/10.13039/501100011033 and FSE+.
This publication has also been funded within the framework of the R\&D\&I Project CEX2025-001574-S, funded by MICIU/AEI/10.13039/501100011033. The research presented in this publication falls within the research line `Origin and Composition of the Universe: Astroparticles and Cosmology (Astro/Cosmo)'.

This work is partially funded by the European Commission – NextGenerationEU, through Momentum CSIC Programme: Develop Your Digital Talent. We acknowledge HPC support by Emilio Ambite, staff hired under the Generation D initiative, promoted by Red.es, an organisation attached to the Spanish Ministry for Digital Transformation and the Civil Service, for the attraction and retention of talent through grants and training contracts, financed by the Recovery, Transformation and Resilience Plan through the EU’s Next Generation funds.

\end{acknowledgments}

\appendix
\section{Differential cross-section for $p+\alpha \rightarrow n + p + \vphantom{}^{3}\rm He$}
\label{app:cross_section_diff}

Throughout this Appendix, we use natural units, i.e., $\hbar=c=1$. 

The differential cross-section for the $p+\alpha \rightarrow n + p + \vphantom{}^{3}\rm He$ process in a general reference frame is derived from the laboratory-frame prescription of Ref.~\cite{Jung1973}:
\begin{equation}
    \frac{d\sigma}{dE_\mathrm{n}} = \frac{C_{^3{ \rm He}}}{p_{p,0}^2}\int_{\chi_{\min}}^{\chi_{max}} e^{-p_{n}^{2}/3\alpha} \chi^{2} \, d\chi,
    \label{eq:dsigma_dE_lab}
\end{equation}
where $p_{p,0}\equiv |\vec{p}_{p,0}|$ and $p_n \equiv |\vec{p}_{n}|$ denote the moduli of the incoming proton and outgoing neutron three-momenta, respectively.\footnote{To avoid confusion about incoming/outgoing particles, in this Appendix the subscript $0$ denotes the incoming particle, while the absence of it denotes the corresponding outgoing particle.} The parameter $\alpha$ is given by $\alpha=0.1455\,\mathrm{MeV}^2$, and $C_{^3\mathrm{He}}=0.592$ is a normalization constant estimated by fitting the theoretical cross-section to available experimental measurements~\cite{Jung1973}. Given the large uncertainties in the data, this value should be regarded as an approximate normalization. 
In this expression, the $\alpha$-particle is assumed to be at rest. 
The integration limits $\chi_{\min}$ and $\chi_{\max}$ are determined by the kinematics of the process. Energy and momentum conservation relate the integration variable $\chi$ to the angle between the incoming proton and outgoing neutron three-momenta through:
\begin{equation}
    \cos{(\vec{p}_{p,0}, \vec{p}_{n})} = \frac{1}{p_{p,0}\,p_{n}}[A p_{p,0}^{2}+ B p_{n}^{2}+C \chi^{2}+ MQ],
    \label{eq:cos_k0_k2}
\end{equation}
where 
\begin{equation}
M=m_p+m_{^3\rm He},
\end{equation}
and
\begin{align}
A &= \frac12\left(1-\frac{M}{m_{p}}\right),\\
B &= \frac12\left(1+\frac{M}{m_n}\right),\\
C\ &= \frac{M}{2\mu}, 
\end{align}
with
\begin{equation}
\mu=\frac{m_p m_{^3\rm He}}{m_p+m_{^3\rm He}}
\end{equation}
the reduced mass of the $p - \, ^3$He system, and $m_p$, $m_n$, and $m_{^3\rm He}$ are the masses at rest of the proton, neutron and $^3\rm He$, respectively. The $Q$ value is taken as
$Q=20.6~{\rm MeV}$~\cite{Jung1973}. 

The condition $ -1 \leq \cos(\vec p_{p,0},\vec p_n) \leq 1 $ implies: 
\begin{equation}
\chi^2=
\frac{\pm \, p_{p,0} \, p_n-Ap_{p,0}^2-Bp_n^2-MQ}{C}.
\end{equation}
The limits $\chi_{\min}$ and $\chi_{\max}$ entering Eq.~\eqref{eq:dsigma_dE_lab} are then obtained from the two solutions above.

\begin{figure}[h!]
    \centering
    \includegraphics[width=1.0\linewidth]{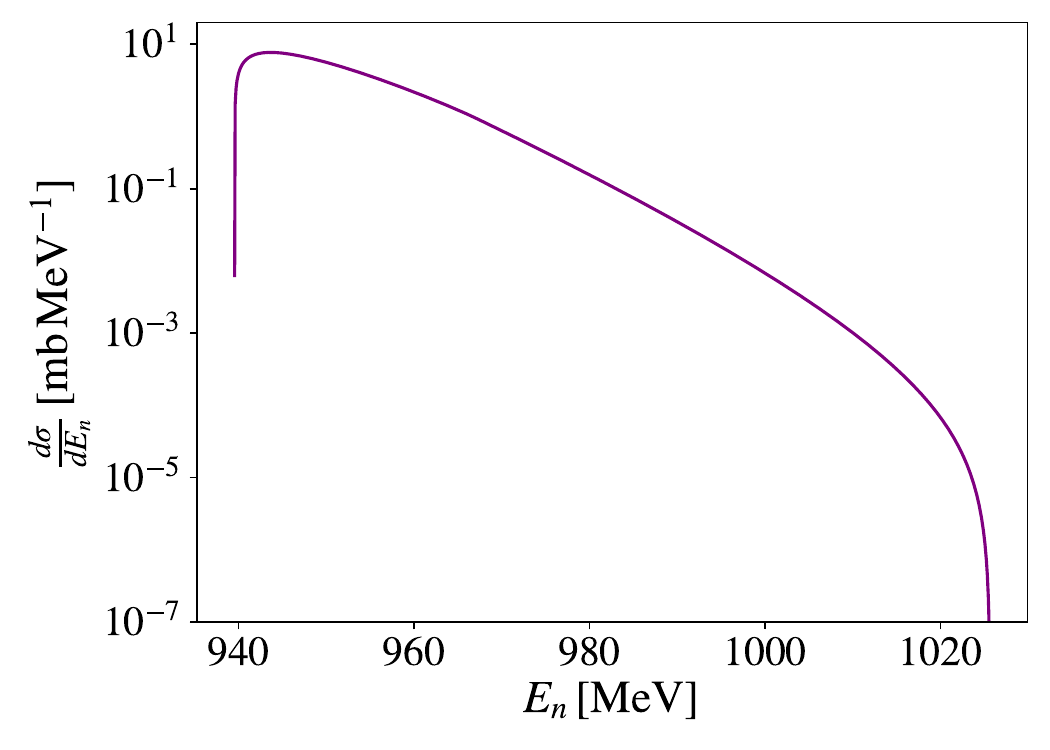}
    \caption{Differential cross-section as a function of the outgoing neutron energy for the $p+\alpha \rightarrow n+p+{}^3$He reaction, assuming an incident proton kinetic energy of $105\,\rm MeV$.}
    \label{fig:dsigma_dEn}
\end{figure}

As an illustrative example, Fig.~\ref{fig:dsigma_dEn} shows the differential cross-section in the laboratory frame as a function of the outgoing neutron energy for an incident proton kinetic energy of $105\,\rm MeV$.

In the accretion flow, both protons and $\alpha$ particles are moving. Therefore, the differential cross-section entering the neutron production rate in Eq.~\eqref{eq:neutron_prduction_rate} must be evaluated in a general reference frame. Starting from the laboratory-frame expression in Eq.~\eqref{eq:dsigma_dE_lab}, we first rewrite the cross-section in fully differential form by replacing the integration variable $\chi$ with the azimuthal ($\phi_n^{\rm lab}$) and polar ($\theta_{\rm pn}^{\rm lab}$) angles of the outgoing neutron with respect to the incoming proton:

\begin{equation}
     \frac{d\sigma^\mathrm{lab}}{dE^\mathrm{lab}_{n} d\cos \theta_{pn}^\mathrm{lab} d\phi_{n}^\mathrm{lab}}  =  \frac{d\sigma^\mathrm{lab}}{dE_{n}^\mathrm{lab} d\chi} \frac{1}{2\pi} \frac{d\chi}{d\cos \theta_{pn}^\mathrm{lab}}.
     \label{eq:sigma_lab_full_angles}
\end{equation}
Differentiating Eq.~\eqref{eq:cos_k0_k2} yields the Jacobian associated with the transformation $\chi\rightarrow\cos(\theta_{pn}^{\rm lab})$,

\begin{equation}
    \frac{d\chi}{d\cos\theta_{pn}^\mathrm{lab}} = \frac{p_{p,0}^\mathrm{lab} p_{n}^\mathrm{lab}}{2 C \chi}.
    \label{eq:angle_tranformation}
\end{equation}

After this change of variables, we express the laboratory-frame differential cross-section in terms of variables defined in a general frame where both the incoming proton and the $\alpha$-particle are moving. Without loss of generality, we choose the $z$-axis to coincide with the direction of motion of the incoming $\alpha$-particle. The neutron energy in this frame is then given by:

\begin{equation}
    E_\mathrm{n}^\mathrm{gen} = \gamma_{\alpha,0}^\mathrm{gen} (E_{n}^\mathrm{lab} + \beta_{\alpha,0}^\mathrm{gen} p_{n}^\mathrm{lab} \cos\theta_{\alpha n}^\mathrm{lab}), 
    \label{eq:lorentz_boost_generic}
\end{equation}
where the superscript $\rm gen$ denotes quantities evaluated in the general reference frame and \begin{equation} \gamma_{\alpha,0}^{\rm gen} = \frac{1}{\sqrt{1-\left(\beta_{\alpha,0}^{\rm gen}\right)^2}}, \end{equation} with $\beta_{\alpha,0}^{\rm gen}$ denoting the velocity of the incoming $\alpha$-particle in that frame.

The angle $\theta_{\alpha n}^{\rm lab}$ is related to the angular variables introduced above through:
\begin{equation} 
\cos \theta_{\alpha n}^{\mathrm {lab }} = \cos \theta_{p \alpha}^{\mathrm {lab }} \cos \theta_{p n}^{\mathrm {lab }} + \sin \theta_{p \alpha}^{\mathrm {lab }} \sin \theta_{p n}^{\mathrm {lab }} \cos \phi_{n}^{\mathrm {lab }}, 
\end{equation} 
where $\theta_{p\alpha}^{\rm lab}$ denotes the angle between the incoming proton and the incoming $\alpha$-particle in the laboratory frame. Applying the Lorentz transformation to the proton momentum, one obtains:

\begin{equation}
\cos \theta_{p \alpha}^{\mathrm {lab }} = \frac{\gamma_{\alpha,0}^{\mathrm {gen }}\left(p_{p,0}^{\mathrm {gen }} \cos \theta_{p \alpha}^{\mathrm {gen }}-\beta_{\alpha,0}^{\mathrm {gen }} E_{p,0}^{\mathrm {gen }}\right)}{p_{p,0}^{\mathrm {lab}}}.
\end{equation}

The energy of the incoming proton in the laboratory frame is related to its value in the general reference frame through:

\begin{equation}
\begin{split}
    E_{p, 0}^{\mathrm {lab }}=\gamma_{\alpha, 0}^{\mathrm {gen }}\left(E_{p, 0}^{\mathrm {gen }}-\beta_{\alpha, 0}^{\mathrm {gen }} p_{p, 0}^{\mathrm {gen }} \cos \theta_{p \alpha}^{\mathrm {gen }}\right),
\end{split}
\end{equation}
\noindent while the proton momentum is $p_{p, 0}^{\mathrm {lab}} = \sqrt{(E_{p, 0}^{\mathrm {lab }})^2 - m_p^2}$.

Since we are interested in the differential cross-section $d\sigma^{\rm gen}/dE_n^{\rm gen}$, the only additional Jacobian required in Eq.~\eqref{eq:sigma_lab_full_angles} is that associated with the transformation between the laboratory-frame and general-frame neutron energies. Keeping $\theta_{pn}^{\rm lab}$ and $\phi_n^{\rm lab}$ fixed, differentiation of Eq.~\eqref{eq:lorentz_boost_generic} yields:

\begin{equation}
    \frac{\partial E_{n}^{\mathrm{gen}}}{\partial E_{n}^{\mathrm{lab}}} = \gamma_{\alpha,0}^\mathrm{gen} \left( 1 + \beta_{\alpha,0}^\mathrm{gen} \frac{E_{n}^{\mathrm{lab}}}{p_{n}^{\mathrm{lab}}} \cos\theta_{\alpha n}^\mathrm{lab} \right)
    \label{eq:energy_transform}
\end{equation}
which allows us to express the laboratory-frame differential cross-section in terms of the neutron energy in the general frame.

Combining Eqs.~\eqref{eq:dsigma_dE_lab}, \eqref{eq:sigma_lab_full_angles}, \eqref{eq:angle_tranformation} and \eqref{eq:energy_transform}, we obtain the differential cross-section in the general frame as:

\begin{equation}
    \begin{split}
    \frac{d\sigma^\mathrm{gen}}{dE_{n}^\mathrm{gen}} = \int_{-1}^1 d\cos{\theta_{pn}^\mathrm{lab}} \int_0^{2\pi} d \phi_{n}^\mathrm{lab} \frac{1}{2\pi} \frac{C_{^3{\rm He}}}{2 C} \frac{p_{n}^\mathrm{lab}}{p_{p,0}^\mathrm{lab}} \\ 
    \times e^{-(p_{n}^\mathrm{lab})^{2}/3\alpha} \chi \left (\frac{\partial E_{n}^{\mathrm{gen}}}{\partial E_{n}^{\mathrm{lab}}}\right )^{-1},
    \end{split}
\end{equation}

\noindent where the laboratory neutron energy in this expression is related to $E_{n}^\mathrm{gen}$ through Eq.~\eqref{eq:lorentz_boost_generic}. 
Note that, after substituting the previous relations, the differential cross-section explicitly depends on the momentum of the incoming proton and $\alpha$-particle ($p_{p,0}^\mathrm{gen}$ and $p_{\alpha,0}^\mathrm{gen}$), as well as on the angle between them, $\theta_{p \alpha}^{\mathrm {gen }}$. These quantities are subsequently integrated over in Eq.~\eqref{eq:neutron_prduction_rate}.

\section{Neutron beta decay kinematics}
\label{app:beta_decay}

In this Appendix, we justify the approximation that the proton produced in neutron beta decay carries essentially all of the energy of the parent neutron. We consider the decay:
\begin{equation}
n \rightarrow p + e^{-} + \bar{\nu}_{e}.
\end{equation}

In the neutron rest frame, denoted by starred quantities, the neutron momentum vanishes ($\vec p_n=0$). Neglecting the antineutrino mass, the total kinetic energy released in the decay is: 
\begin{equation}
Q=m_n-m_p-m_e\simeq 0.782~{\rm MeV},
\end{equation}
where $m_n=939.565 \, \rm MeV$, $m_p=938.272 \, \rm MeV$, and $m_e= 0.511 \, \rm MeV$.

Energy conservation then gives:
\begin{equation}
m_n=E_p^{*}+E_e^{*}+E_{\bar{\nu_e}}^{*}.
\end{equation}

\noindent Equivalently, in terms of the kinetic energies:
\begin{equation}
Q=K_p^{*}+K_e^{*}+E_{\bar{\nu}_e}^{*},
\end{equation}
where $E_{\bar{\nu}_e}^{*}=K_{\bar{\nu}_e}^{*}$, since the antineutrino mass is neglected.

Momentum conservation additionally requires
\begin{equation}
\vec{p}_p^{\,*}=-\left(\vec{p}_e^{\, *}+\vec{p}_{\bar{\nu}_e}^{\, *}
\right).
\end{equation}

\noindent The modulus of the electron momentum satisfies
\begin{equation}
p_e^{*}=\sqrt{\left(K_{e}^{*}\right)^2+2m_eK_e^{*}},
\end{equation}
and the modulus of the neutrino momentum $p_{\bar{\nu}_e}^{*}=E_{\bar{\nu}_e}^{*}$ 

\noindent Since the total kinetic energy released in the decay is  $Q\simeq0.782~{\rm MeV}$, all decay products carry momenta of order MeV or smaller. In particular, momentum conservation implies that the proton recoil momentum magnitude satisfies $p_p^*=\mathcal O({\rm MeV})$. 
Since $p_p^* \ll m_p$, the proton is non-relativistic in the neutron rest frame. Its kinetic energy is therefore well approximated by:
\begin{equation}
K_p^{*}=E_p^{*}-m_p\simeq\frac{(p_p^{*})^2}{2m_p}.
\end{equation}

The maximum proton recoil momentum is reached at the endpoint of the electron spectrum, where the antineutrino carries vanishingly small energy. In this limit:

\begin{equation}
E_{e,\max}^{*}=
\frac{m_n^2+m_e^2-m_p^2}{2m_n},
\end{equation}
and therefore:
\begin{equation}
p_{p,\max}^{*}=
p_{e,\max}^{*}=
\sqrt{\left(E_{e,\max}^{*}\right)^2-m_e^2}
\simeq1.19~{\rm MeV}.
\end{equation}

Since the proton is non-relativistic in the neutron rest frame, its maximum recoil kinetic energy is:
\begin{equation}
K_{p,\max}^{*}
\simeq
\frac{\left(p_{p,\max}^{*}\right)^2}{2m_p}
\simeq0.75~{\rm keV}.
\end{equation}
Thus, to a good approximation,
\begin{equation}
p_p^{*}\lesssim 1.19 \, \mathrm{MeV},
\qquad
K_p^{*}\lesssim 0.75 \, \rm keV.
\end{equation}
Hence, the electron and antineutrino carry essentially all of the available decay energy, while the proton remains nearly at rest in the neutron rest frame.

Since the decay occurs over microscopic length scales, its kinematics can be described in a local inertial frame. In the local plasma frame, the parent neutron moves with velocity $\beta_n$ and the Lorentz factor is:
\begin{equation}
\gamma_n=\frac{E_n}{m_n}=1+\frac{K_n}{m_n},
\end{equation}
where $K_n=E_n-m_n$ is the neutron kinetic energy in this frame. The energy of the proton immediately after the decay is:
\begin{equation}
E_p=\gamma_n\left(E_p^{*}+\beta_n p_p^{*}\cos\theta^{*}\right),
\end{equation}
where $\theta^*$ is the angle between the proton momentum in the neutron rest frame and the boost direction. The energy of the parent neutron in the same local frame is:
\begin{equation}
E_n=\gamma_n m_n.
\end{equation}

The ratio of the two energies is then:
\begin{equation}
\frac{E_p}{E_n}=\frac{E_p^{*}}{m_n}+
\beta_n\frac{p_p^{*}}{m_n}\cos\theta^{*}.
\end{equation}
Using $E_p^{*}=m_p+K_p^{*}$, this becomes:
\begin{equation}
\frac{E_p}{E_n}=\frac{m_p}{m_n}+\frac{K_p^{*}}{m_n}+\beta_n\frac{p_p^{*}}{m_n}\cos\theta^{*}.
\end{equation}
The terms governing the difference between the proton and neutron energies satisfy:
\begin{equation}
\frac{m_n-m_p}{m_n}
\simeq1.4\times10^{-3},
\hspace{0.25cm}
\frac{K_p^{*}}{m_n}
\lesssim8\times10^{-7},
\hspace{0.25cm}
\frac{p_p^{*}}{m_n}
\lesssim1.3\times10^{-3}.
\end{equation}
Consequently,
\begin{equation}
\frac{E_p}{E_n}
=
1+\mathcal{O}(10^{-3}),
\end{equation}
and therefore: 
\begin{equation}
E_p\simeq E_n.
\end{equation}

The proton produced in neutron beta decay inherits the energy of the parent neutron up to sub-percent corrections. Neglecting these corrections is therefore fully justified for the purposes of the present work.

\bibliography{apssamp}

\end{document}